%% file: SNTemplates.tex
\documentclass[aps,prf,nofootinbib,twocolumn,showkeys,showpacs]{revtex4}

\usepackage{amsmath}
\usepackage{amssymb}
\usepackage[utf8]{inputenc}
\usepackage{caption}
\usepackage{graphicx}
\usepackage{multirow}
\usepackage{xcolor}
\usepackage{color}
\usepackage{aas_macros}
\usepackage{ulem} 
\usepackage{subfig}
\usepackage[export]{adjustbox}
\usepackage{enumerate}

\begin{document}
\title{Phenomenological gravitational waveforms for core-collapse supernovae}
%\author{Authors}
\author{P. Cerd\'{a}-Dur\'{a}n$^{1,2}$,
M. L\'{o}pez $^{3,4}$, 
A. Favali $^{5}$
I. Di Palma$^{5,6}$, 
M. Drago$^{5,6}$, 
F. Ricci$^{5,6}$}

\affiliation{$^1$ Departamento de Astronom\'ia y Astrof\'isica, Universitat de Val\`encia, Dr. Moliner 50, 46100, Burjassot (Valencia), Spain}
\affiliation{$^2$ Observatori Astron\`omic, Universitat de Val\`encia, Catedr\'atico Jos\'e Beltr\'an 2, 46980, Paterna, Spain}
\affiliation{$^3$ Institute for Gravitational and Subatomic Physics (GRASP), 
Department of Physics, Utrecht University, 
Princetonplein 1, 3584 CC Utrecht, The Netherlands}
\affiliation{$^{4}$ Nikhef, Science Park 105, 1098 XG Amsterdam, The Netherlands}
\affiliation{$^{5}$ Universit\`a di Roma  {\it{La Sapienza}}, I-00185 Roma, Italy}
\affiliation{$^{6}$ INFN, Sezione di Roma, I-00185 Roma, Italy}
%\affiliation{$^{7}$ Gran Sasso Science Institute (GSSI), I-67100 L'Aquila, Italy}
%\affiliation{$^{8}$ INFN, Laboratori Nazionali del Gran Sasso, I-67100 Assergi, Italy}
%\affiliation{$^{7}$ Department of Physics, Ariel University, Ariel, Israel}

%\date{\today}
\begin{abstract}
Galactic core-collapse supernovae (CCSNe) are a target for current generation gravitational wave detectors with an expected rate of 1-3 per century. The development of data analysis methods used for their detection relies deeply on the availability of waveform templates. However, realistic numerical simulations producing such waveforms are computationally expensive (millions of CPU hours and $10^2-10^3$~GB of memory), and only a few tens of them are available nowadays in the literature. We have developed a novel parametrized phenomenological waveform generator for CCSNe, \texttt{ccphen v4}, that reproduces the morphology of numerical simulation waveforms with low computational cost ($\sim 10$~ms CPU time and a few MB of memory use).  For the first time, the phenomenological waveforms include polarization and the effect of several oscillation modes in the proto-neutron star. This is sufficient to describe the case of non-rotating progenitor cores, representing the vast majority of possible events. The waveforms include a stochastic component and are calibrated using numerical simulation data. The code is publicly available. Their main application is the training of neural networks used in detection pipelines, but other applications in this context are also discussed. 
 \end{abstract}
\keywords{Gravitational waves, Core collapse supernovae, Astrophysics.}

\maketitle

%\idp{insert the orcid numbers for the authors}

\input{introduction}
\input{oscillations}
\input{parameters}

\input{ccphen}
\input{comparison}
\input{conclusions}

\vskip 2mm
\section{Acknowledgement}
PCD acknowledges the support from the grants 
Prometeo CIPROM/2022/49 from the Generalitat Valenciana, and  PID2021-125485NB-C21 from the Spanish Agencia Estatal de Investigación funded by MCIN/AEI/10.13039/501100011033 and ERDF A way of making Europe. IDP acknowledges the support of the  Sapienza  School for Advanced Studies (SSAS) and the support of the Sapienza Grants No. RM120172AEF49A82 and RG12117A87956C66. MD acknowledges the support of the Sapienza Grants No. RM123188F3F2172A and RM1221816813FFA3. ML is supported by the research program of the Netherlands Organization for Scientific Research (NWO). The authors are grateful for computational resources provided by the LIGO Laboratory and supported by the National Science Foundation Grants No. PHY-0757058 and No. PHY-0823459. 

\appendix
\input{appendix}

%%%%%%%%%%%%%%%%%%%%%%%%%%%%%%%%%%%%%%%%%%%%%%%%%%%%%%%%%%%%%%%%%
%%%%%%%%%%%%%%%%%%%%%%%%%%%%%%%%%%%%%%%%%%%%%%%%%%%%%%%%%%%%%%%%%
%%%%%%%%%%%%%%%%%%%%%%%%%%%%%%%%%%%%%%%%%%%%%%%%%%%%%%%%%%%%%%%%%

\bibliographystyle{unsrt}
\bibliography{biblio}

%\printbibliography
%\bibliography{references}
%\end{multicols}

\end{document}

%% file: introduction.tex
%%%%%%%%%%%%%%%%%%%%%%%%%%%%%%%%%%%%%%%%%%%%%%%%%%%%%%%%%%%%%%%%%
%%%%%%%%%%%%%%%%%%%%%%%%%%%%%%%%%%%%%%%%%%%%%%%%%%%%%%%%%%%%%%%%%
\section{Introduction}
\label{sec:introduction}

Since the first detection of a binary black hole merger in 2015, which established the field of gravitational wave (GW) {astronomy, about a hundred detections have been reported} \cite{LIGOScientific:2016aoc}. In 2017 the first observation of a binary neutron star merger by the Advanced LIGO \cite{LIGOScientific:2014pky} and Advanced Virgo \cite{VIRGO:2014yos} {detectors, coincident with the electromagnetic detection of a short gamma-ray burst and a kilonova \cite{Goldstein:2017mmi, LIGOScientific:2017ync}} initiated the era of multi-messenger astronomy (MMA) with GW: a novel paradigm capable of probing the densest and most energetic regions of the universe. 

%and joint observation of Advanced LIGO \cite{LIGOScientific:2014pky} and Advanced Virgo \cite{VIRGO:2014yos} with Fermi Gamma-ray Burst Monitor \cite{Meegan:2009qu} initiated the era of multi-messenger astronomy (MMA) with GW: a novel paradigm capable of probing the densest and most energetic regions of the universe \cite{Goldstein:2017mmi, LIGOScientific:2017ync}.

The improvement of GW detectors has led to the successful discovery of tens of compact binary coalescence (CBC) events \cite{LIGOScientific:2018mvr, LIGOScientific:2020ibl, LIGOScientific:2021djp}. Nonetheless, {the} GW emission {associated to} core-collapse supernova (CCSN) {explosions}, another important target for MMA, has not yet been observed due to the weakness of the signal and its inherent complexity: despite being one of the most energetic events in the universe, the expected {strain} amplitude of {a} Galactic CCSN ranges $\sim 10^{-21} - 10^{-23}$, with {a rate of $\sim 1 - 3$} per century {(see \cite{Gossan:2016,Abbott:2020} and references therein).}

At the end of their lives, massive stars {of} $\sim 8 - 100\,$M$_{\odot}$ have accumulated $\approx 1.4 $ M$_{\odot}$ of elements of the iron family in a compact core due to thermonuclear fusion processes. {Having reached the Chandrasekhar mass,} the iron core cannot support its own weight and {undergoes}
a gravitational collapse. {As the density increases up to nuclear saturation density, heavy nuclei are disintegrated into free nucleons, producing neutrinos that become trapped. At the same time, the} sharp rise of the incompressibility, due to repulsive contributions to the nuclear force between the nucleons, halts the collapse of the inner core forming a proto-neutron star (PNS). As it bounces back, it produces a shock wave that stalls at about 100 km. Helped by the additional thermal energy deposited by the neutrinos {diffusing out of the PNS}, the shock may revive in timescales of hundreds of milliseconds disrupting the entire star and producing an electromagnetic signal known as supernova \cite{RevModPhys.62.801, Janka:2017vlw}.

Mass motions in the newly formed PNS are responsible for the emission of strong GW signals, which could be detectable at galactic distances \cite{Szczepanczyk2021}. The basic theory of CCSN explosion is consistent with SN 1987 A, which was detected via electromagnetic radiation and low energy neutrino emission \cite{PhysRevLett.58.2722}. Nonetheless, GW and neutrino emission, as opposed to electromagnetic radiation, provide direct and unique information about the inner dynamics of the collapse, since they are produced at its inner core. Thus, combined multi-messenger detection of the GW signal together with the neutrino signal would be critical to confirm this theoretical model, and would help us understand the details of the processes taking place during the explosion.

In the context of GW searches, the template-matching techniques \cite{Usman:2015kfa} that have  {been successfully used in the detection of CBC signals}, are 
not feasible for the case of CCSNe. The main reason is that the signals are intrinsically stochastic because the emission process is tightly related to the development of instabilities in the PNS (e.g. convection). Therefore, waveforms from numerical simulations should be regarded as particular realizations of all possible signals that a system with the same features (progenitor mass, rotation, ...) may produce,  rather than the true template for this system. 

Additionally, the generation of large template banks of CCSN waveforms is challenging due to the large computational cost of numerical simulations.
%unfeasible since the generation of numerical simulations is challenging and computationally intensive.
{Therefore}, current CCSN searches, {both} targeted \cite{LIGOScientific:2016jvu, LIGOScientific:2019ryq, Szczepanczyk:2023ihe} {and} all-sky \cite{LIGOScientific:2016kum, LIGOScientific:2019ryq, Szczepanczyk:2023ihe}, employ model-free algorithms that rely on excess of power to identify signals buried in detector noise, such as coherent Waveburst \cite{Drago:2020kic} %, oLIB \cite{Lynch:2015yin} 
or BayesWave \cite{Cornish:2014kda}. These state-of-the-art algorithms reach $50 \%$ detection efficiency at 5 kpc during O1 and O2, at 8.9 kpc during O3, and could {reach} $\sim 10$ kpc during the advanced detector era{, for the most common case of slow-rotating progenitors}. To improve their detection probability we have to broaden the volume of the universe to be explored, which can be achieved by improving the detectors' sensitivity or designing robust and novel methods to enhance CCSN searches.
 
 In this line of thought, several authors have proposed to benefit of the particularity of CCSN signal for detection \cite{Astone:2018uge, Chan:2019fuz, Cavaglia:2020qzp, Iess:2020yqj, Lopez2021, Lopez:2021rci, Mukherjee:2021xdt, Iess:2023quq, 2023Mitra:stad169}, and regression \cite{Edwards:2020hmd, Lagos:2023qli} tasks using Machine Learning (ML). ML has been successful in a variety of applications and it has emerged as a novel tool in the GW field (see \cite{Cuoco:2020ogp} for a comprehensive review). These methods are able to perform analysis rapidly since all the intensive computation is diverted to the one-time training stage. Due to their ability to grasp the underlying characteristics of the data, it is preferable to use large data sets to cover as deep a parameter space as possible.

Nevertheless, because of the massive amount of computational resources needed to produce each of {the} CCSN simulations, there is a limited number of them. This imposes a challenge for ML applications due to the lack of data: {as we discuss in detail in the next sections, most realistic 3D simulations amount to a few tens, and even including less intensive (and less accurate) simulations, there are no more than $\sim 100$ available in the literature. Some authors \cite{richers2017} have computed several thousand simulations, but they correspond only to the signal at bounce, which is a very small fraction of the whole signal (only the first $\sim 25$ ms).}
 Furthermore, due to their numerical challenge and the various approximations used, it is unclear how close the existing numerical waveforms correspond to the actual GW signal for a specific type of progenitor {or what is the dependence of the waveform properties with the progenitor properties (there is only partial coverage of the CCSN parameter space). }
 % since they just have partial coverage of the CCSN parameter space.

 To overcome this issue {\cite{Astone:2018uge} and \cite{Lopez2021} developed a set of phenomenological waveforms with a stochastic component that are inexpensive to generate and mimic in time-frequency the rising arch of the dominant mode of CCSN signal. These waveforms have been used to train convolutional neural networks (CNNs) that are able to detect \cite{Astone:2018uge, Lopez2021} or infer the properties \cite{Lagos:2023qli} of realistic CCSN waveforms from numerical simulations, proving that this is a valid approach and that phenomenological waveforms are sufficiently close to realistic ones to serve as a base for training CNNs. The phenomenological waveforms of \cite{Astone:2018uge, Lopez2021} have several limitations:

 \begin{enumerate}[i)]
 \item They do not include polarization, so they can be used as a prediction of the strain at one single detector, but cannot be used properly for networks of detectors.
 
 \item They only contain the contribution of the dominant mode in the signal but not additional modes that are expected to appear (see discussion below).
 
 \item The parameter space and calibration used was based on a limited set of simulations and applied to a limited set of parameters being, for many parameters, just best guesses. 
 \end{enumerate}
 
 A similar approach has been followed by \cite{Powell2022,Powell2023} which used frequency-varying chirplets as phenomenological templates. Although in this case polarization is considered, the waveforms are much more simplified than in \cite{Astone:2018uge, Lopez2021} and do not contain any stochastic component.}
 In this work we present a new phenomenological waveform generator that extends the work by \cite{Astone:2018uge, Lopez2021} addressing the three aforementioned issues.
 
 The structure of this paper is as follows: in Sect.~\ref{sec:PNSoscillations} we describe the current understanding of PNS oscillations as the dominant GW emission mechanism in CCSN and present a simplified model for these oscillations that can be used to build phenomenological waveforms. In Sect.~\ref{sec:Parameters} we investigate the space of parameters of the phenomenological waveforms by analyzing a set of realistic numerical simulations available in the literature. In Sect.~\ref{sec:ccphen} we present the details of the numerical code computing the phenomenological waveforms, some examples of waveforms and code tests. In Sect.~\ref{sec:comparison} we compare the new phenomenological waveforms directly with realistic CCSN waveforms to assess their similarity. Finally, in Sect.~\ref{sec:Conclusions} we present our conclusions.

%% file: oscillations.tex
\section{Proto-neutron star oscillations}
\label{sec:PNSoscillations}

For non-rotating stars, we can regard the PNS as a spherically symmetric object whose deviations from
spherical symmetry can be modelled as perturbations. After the bounce, the properties of the background spherical object 
change slowly with time, in timescales of $\sim 100$~ms set by the PNS cooling and the accretion timescales. This timescale is much smaller than the timescale associated with most of the perturbations, therefore, at any given time, the perturbations can be modelled as the solution of an eigenvalue problem resulting from the linearization of the hydrodynamics equations around the hydrostatic equilibrium. The result of the eigenvalue analysis is the existence of oscillatory modes in the form of (buoyancy-driven) g-modes, (pressure-driven) p-modes and f-modes. This approach has proven to describe the behaviour of the PNS in the post-bounce evolution and is able to explain the main features in the GW spectrum of CCSNe \citep{Torres-Forne2018,Morozova2018,Torres-Forne2019a,Sotani2019,Westernacher-Schneider2019,Westernacher-Schneider2020,Sotani2020}. When considered the accretion flow into the PNS, additional modes appear associated to the acoustic-advective interaction between the shock and the PNS, in a process known as the standing accretion shock instability (SASI) \cite{Foglizzo2000,Blondin2003}. 

We aim to build a simplified model for this scenario based on the existence of PNS modes emitting GWs. With this in mind, we proceed next to show the framework in which our GWs are computed, to present a simple model for the evolution of an initially perturbed PNS, and finally build a model for a continuously excited PNS. The model presented here is an extension of the work of \cite{Astone2018, Lopez2021} that allows the computation of both polarizations of the GW signal.

%%%%%%%%%%%%%%%%%
\subsection{Quadupolar approximation of the GW emission}

In the slow-motion approximation, the GW emission can be approximated by the Einstein quadrupole formula \citep[see e.g.][]{Thorne1980}. In the case of core-collapse supernovae, this approximation reproduces with high accuracy the phase of the gravitational wave signal and its amplitude within a $10\%$ error \cite{Reisswig2011}. For this reason, it is used regularly in core-collapse simulations to compute the GW signal (in particular all the simulations mentioned in this work).

The strain of the GW emission in the quadrupolar approximation can be  expressed in a very compact form in terms of the spin-weighted spherical harmonics with $s=-2$
\citep[see e.g.][]{Thorne1980}:
\begin{align}
h = h_+-i h_\times& =  \frac{1}{D} \frac{G}{c^4}\frac{8\pi}{5} \sqrt{\frac{2}{3}} \sum_{m=-2}^{+2} \ddot I_{2m} \,\, {}^{}_{-2}Y^{2m}(\Theta,\Phi),
\label{eq:quadrupoleformula}
\end{align}
where $D$ is the distance to the source. $I_{lm}$ are a complex quantities, the multipolar moments, defined as 
\begin{equation}
    I_{lm} \equiv  \int \rho ({\bf x}, t) \, r^l Y_{lm}^*(\theta,\varphi)  d^3{\bf x},
    \label{e:Ilm}
\end{equation}
where $\rho$ is the rest-mass density. The multipolar moments with the different sign of $m$ fulfil
\begin{equation}
I_{lm} = (-1)^m I^*_{l-m}, 
\end{equation}
For $l=2$, we refer to it as the mass quadrupole moment.
The values of the $s=-2$ spin-weighted tensor spherical harmonics with $l=2$ are:
\begin{align}
{}^{}_{-2}Y^{20}(\Theta,\Phi) &= \frac{1}{4}\sqrt{\frac{15}{2\pi}}\sin^2\Theta, \\
{}^{}_{-2}Y^{2\pm1}(\Theta,\Phi) &= \frac{1}{8}\sqrt{\frac{5}{\pi}}  (2 \sin\Theta \pm \sin 2\Theta) e^{\pm i\,\Phi}, \\
{}^{}_{-2}Y^{2\pm2}(\Theta,\Phi) &= \frac{1}{16}\sqrt{\frac{5}{\pi}} (3 \pm 4 \cos\Theta +\cos 2\Theta) e^{\pm i\,2\Phi}.
\end{align}
To avoid confusion, we use  $\theta$ and $\varphi$ for the angles of the spherical coordinates centred in the source when used for the integration 
of the mass quadrupole, and $\Theta$ and $\Phi$ for similar angles when used to denote the observation angle. 
These tensor spherical harmonics fulfill the orthonormality relation
\begin{equation}
\int d\Omega {}_{s}Y^{lm} {}_{s}Y^{l'm'*} = \delta_{ll'} \delta_{mm'}. \label{eq:YlmOrtho}
\end{equation}
This allows us to express the strain as 
\begin{equation}
h = \sum_{m=-2}^{m=2} h_{2m} \, {}^{}_{-2}Y^{2m}(\Theta,\Phi)
\end{equation}
where
\begin{equation}
h_{2m} = \frac{1}{D} \frac{G}{c^4}\frac{8\pi}{5} \sqrt{\frac{2}{3}} \ddot I_{2m}.
\end{equation}

We define the root-mean-square of the angular average of $h$ as 
\begin{eqnarray}
\bar h_{\rm rms} &\equiv& \sqrt{\left < \, \frac{1}{4\pi}\int d\Omega \, |h|^2 \right>} \nonumber\\
&=& \sqrt{ \frac{1}{4\pi} \sum_{m=-2}^{+2}  \left<|h_{2m}|^2 \right >}, 
\label{eq:hrms}
\end{eqnarray}
where we have used the orthonormality relations of Eq.~\eqref{eq:YlmOrtho} and
\begin{equation}
\left< f \right > \equiv \frac{1}{T} \int_0^T dt f,
\label{eq:timeAverage}
\end{equation}
is the time average.

We consider statistically isotropic sources. This is valid for the collapse of non-rotating progenitors, which are approximately spherically symmetric. This symmetry is broken after the collapse but, in a statistical sense, it is still spherically symmetric. The reason is that if we consider a sufficiently large ensemble of realizations of the collapse of the same progenitor core, the outcome averaged over all realizations is spherically symmetric. Under this assumption, it holds that
\begin{equation}
 <|I_{20}|^2> = <|I_{21}|^2> = <|I_{22}|^2>.
\end{equation}
A proof for these conditions can be found in Appendix~\ref{app:isotropy}. These conditions are used in our waveform generator to impose statistical isotropy on the source,
as described in the next sections, by fixing the relative rms amplitude of $I_{2m}$.

%%%%%%%%%%%%%%%%%
\subsection{Oscillations of an initially perturbed static proto-neutron star}
\label{sec:oscillations}

Let us consider first a PNS described by a spherically symmetric static background with density $\rho_0(r)$. This background admits oscillatory solutions (modes), a solution of an eigenvalue problem for the linearly perturbed equations expanded in terms of the spherical harmonics $Y_{lm}(\theta,\varphi)$. The real and imaginary parts of the resulting eigenvalues, $\omega^{(T)}_{l}$ and $\sigma^{(T)}_{l}$ depend on the number $l$ and of the type of mode (e.g. g-modes, p-modes, f-modes, SASI ..., noted here with the label $(T)$). Note that only modes with $l= 2$ contribute to the dominant quadrupolar GW signal and modes with different values of $m$ have the same frequency and damping rate for spherically symmetric backgrounds.  

If we consider an initially perturbed PNS, we can describe the time evolution of the rest-mass density as
\begin{align}
&\rho ({\bf x}, t) = \rho_0 (r) \nonumber \\
&+ \mathcal{R}\left \{ \sum_{T} \sum_{l=0}^{\infty} \sum_{m = -l}^{+l} \rho^{(T)}_{lm} (r) \, Y_{lm} (\theta,\varphi) \, e^{i \left (\omega^{(T)}_{l} - i \sigma^{(T)}_{l} \right) t} \right \},
\end{align}
where  $\rho^{(T)}_{lm}$ is the amplitude of a particular oscillation mode and $\mathcal{R}$ indicates the real part. This expression is valid for a single initial perturbation, that sets the value of $\rho^{(T)}_{lm}$, and results in a combined damped oscillation of the different modes, decaying exponentially in time towards the background.

In the next derivations, we focus on a single eigenmode. Therefore, to simplify the notation we remove the label $(T)$, unless strictly necessary. The time-varying part of the quadrupolar moment can be therefore computed as
\begin{align}
&I'_{lm} \equiv  \mathcal{R} \left \{ \int (\rho ({\bf x}, t) - \rho_0(r) ) r^l Y_{lm}^*(\theta,\varphi)  d^3{\bf x} \right \}\nonumber\\
&=\frac{1}{2} \left [ A_{lm} \, e^{i  (\omega_{l} + i \sigma_{l} ) t} + (-1)^m A_{l-m} \, e^{-i  (\omega_{l} + i \sigma_{l} ) t} \right ],
\label{eq:Ilmprime}
\end{align}
where
\begin{equation}
A_{lm} \equiv \int \rho_{lm} (r)\,   r^{(l+2)} dr.
\end{equation}
Since the density perturbations $\rho_{lm}$ are real functions but $I'_{lm}$ is a complex function, the relation between both mixes inevitably positive and negative $m$. Note that $\ddot I_{lm} = \ddot I'_{lm}$, so it can be used directly in the quadrupole formula. 
This quantity fulfils the equation for the damped harmonic oscillator
\begin{equation}
\ddot I'_{lm} - 2 \sigma \dot I'_{lm} + (\omega^2 + \sigma^2 ) I'_{lm} = 0. 
\end{equation}
Using the natural frequency $\omega_0 \equiv \sqrt{\omega^2 + \sigma^2}$ 
and the quality factor $Q \equiv -\omega_0/(2\sigma)$, instead of $\omega$ and $\sigma$, we can rewrite this equation as
\begin{equation}
\ddot I'_{lm} + \frac{\omega_0}{Q} \dot I'_{lm} + \omega_0^2  I'_{lm} = 0. \label{eq:HarmonicOscllator}
\end{equation}
Since $I'_{lm}$ are complex quantities, for a given $l$, we would have to integrate $2 l+1$ complex quantities, or equivalently $2(2 l +1)$ real quantities. However, these quantities fulfil that
\begin{equation}
I'_{lm} = (-1)^m I'^*_{l-m}, 
\end{equation}
or equivalently, splitting into real and imaginary part
\begin{eqnarray}
\mathcal{R}(I'_{lm}) =& (-1)^m & \mathcal{R} (I'_{l-m}),\nonumber \\
\mathcal{I}(I'_{lm}) =& (-1)^{m+1}&  \mathcal{I} (I'_{l-m}), 
\end{eqnarray}
where $\mathcal{I}$ denotes the imaginary part. These relations imply that all moments with negative $m$ can be expressed in terms of moments with positive $m$ and that $\mathcal{I}(I'_{l0})=0$. In practice it means that we just have to integrate $2l+1$ real quantities and the rest can be computed from those, i.e. the number of degrees of freedom is the same as for $\rho_{lm}$, as one should expect. 

For the case considered here, with $l=2$, this means that we just integrate $\mathcal{R}(I'_{20})$, $\mathcal{R}(I'_{21})$, $\mathcal{R}(I'_{22})$, $\mathcal{I}(I'_{21})$, $\mathcal{I}(I'_{22})$, and then compute the others as
\begin{align}
    \mathcal{I}(I'_{20}) &= 0, & \nonumber \\
    \mathcal{R}(I'_{2-1}) &= -\mathcal{R}(I'_{2,1}), & 
    \mathcal{R}(I'_{2-2}) &= \mathcal{R}(I'_{2,2}), \nonumber \\
    \mathcal{I}(I'_{2-1}) &= \mathcal{I}(I'_{2,1}), &
    \mathcal{I}(I'_{2-2}) &= -\mathcal{I}(I'_{2,2}). 
\end{align}
%

%%%%%%%%%%%%%%%%%
\subsection{Oscillations of a continuously-excited slowly-evolving proto-neutron star}

In a realistic model of CCSNe the PNS, oscillation modes are excited continuously by the action of non-linearities arising from the effect of several instabilities, including neutrino-driven convection and SASI, and from external perturbations of an accreting fluid that, in general, may be anisotropic. Additionally, the slowly evolving background modifies continuously the eigenvalues of the system. A full predictive analysis of the oscillation mode evolution would require performing numerical simulations and then computing the time-varying values of the eigenvalues \citep[This would be the approach followed by][]{Torres-Forne2018,Morozova2018,Torres-Forne2019a,Sotani2019,Sotani2020}. However, our phenomenological approach to waveform generation aims to avoid the computationally expensive numerical simulations of CCSNs. Following this philosophy, we consider a generic $\omega(t)$ and $\sigma(t)$ that represents the evolution of the real and imaginary part of the eigenvalue of a mode, respectively, or equivalently $\omega_0(t)$ and $Q(t)$.

Therefore, the natural extension of Eq.~\eqref{eq:HarmonicOscllator} to the case of a continuously-excited slowly-evolving PNS is a driven damped harmonic oscillator of the form
\begin{equation}
\ddot I'_{lm} + \frac{\omega_0(t)}{Q(t)} \dot I'_{lm} + \omega_0(t)^2  I'_{lm} = W_{lm}(t),
\label{eq:harmonic}
\end{equation}
where the driving term $W_{lm}(t)$ has units of energy and is related to the work done by the external forcing (the perturbations mentioned above) on the system. Note that the forcing is independent for each of the components $lm$. If we define a driving force density ${\bf f}$ acting on a fluid element to displace it by $d{\bf l}$, the differential work per unit volume is
\begin{equation}
\delta w={\bf f}\cdot d{\bf l} =  {\bf f}\cdot {\bf v} \, dt = p \,dt,
\end{equation}
where $p\equiv{\bf f}\cdot {\bf v}$ is a power density. Integrating over the whole volume of the system
\begin{equation}
\delta W= \int d^3{\bf x} \, {\bf f}\cdot {\bf v} \, dt = \int d^3{\bf x} \,  p \,dt =P \,dt,
\end{equation}
where $P$ is the power related to the external forcing. The work done at any time (now adding the $lm$ label) is just given by
\begin{equation}
W_{lm}(t) = \int_0^t P_{lm}(t') dt'
\label{eq:power}
\end{equation}

The numerical time integration of Eq.~\eqref{eq:quadrupoleformula} together with Eq.~\eqref{eq:harmonic} can be used to generate GW signals similar to those found in CCSN simulations, provided that realistic values for $\omega_0(t)$, $Q(t)$ and $W_{lm}(t)$ are given. We address the choice of all relevant parameters for the waveform in the next section.

%% file: parameters.tex
%%%%%%%%%%%%%%%%%%%%%%%%%%%%%%%%%%%%%%%%%%%%%%%%%%%%%%%%%%%%%%%%%
%%%%%%%%%%%%%%%%%%%%%%%%%%%%%%%%%%%%%%%%%%%%%%%%%%%%%%%%%%%%%%%%%
\section{Parameters choice and calibration}
\label{sec:Parameters}

\begin{table*}
 \centering 
 \caption{List of 3D CCSN simulations used to calibrate the parameters of our phenomenological waveforms. See the text for a description of the labels. }
 \label{tab:CCSNModels}
 \begin{tabular}{llccllc}
  \hline\hline
   Source & $\qquad$ &  Neutrino transport & Gravity &Model & EOS & $M_{\rm ZAMS}$  \\ \hline
   Kuroda et al. (2016) &\cite{Kuroda:2016} & M1 & BSSN& SFHx & SFHx & $15.0$ \\
   &&&& TM1 & TM1 & $15.0$ \\ 
   \hline
   Andresen et al. (2017) &\cite{Andresen:2017} & Ray-by-ray+ & TOV & s11.2 & LS220 & $11.0$  \\
   &&&& s20 & LS220 & $20.0$\\
   &&&& s20s & LS220 & $20.0$\\
   &&&& s27 & LS220 & $27.0$ \\
   \hline
   Kuroda et al. (2017) & \cite{Kuroda2017} & M1 & BSSN & s11.2 & SFHx & $11.2$ \\
   &&&& s15.0 & SFHx & $15.0$ \\
   \hline
   Yakunin et al. (2017) & \cite{Yakunin2017} & MGFLD & TOV & C15-3D & LS220 & $15.0$ \\
   \hline
   O'Connor \& Couch (2018) & \cite{O'Connor2018} & M1 & TOV  & mesa20 & SFHo & $20.0$\\
   &&&& mesa20\_pert & SFHo & $20.0$ \\
   &&&& mesa20\_v\_LR & SFHo & $20.0$ \\
   &&&& mesa20\_pert\_LR & SFHo & $20.0$ \\
   &&&& mesa20\_LR & SFHo & $20.0$ \\
   \hline
   Andresen et al (2019) & \cite{Andresen2019} & Ray-by-ray+ & TOV & m15nr & LS220 & $15.0$ \\
   \hline
   Powell \& M\"uller (2019) & \cite{Powell2019} & FMT & XCFC & s18 & LS220 & $18.0$ \\
   \hline
   Radice et al. (2019) & \cite{Radice2019} & MGFLD & TOV & s9 & SFHo & $9.0$ \\
   &&&& s10 & SFHo & $10.0$ \\
   &&&& s11 & SFHo & $11.0$ \\
   &&&& s12 & SFHo & $12.0$ \\
   &&&& s13 & SFHo & $13.0$ \\
   &&&& s19 & SFHo & $19.0$ \\
   &&&& s25 & SFHo & $25.0$ \\
   &&&& s60 & SFHo & $60.0$ \\
   \hline
   Powell \& M\"uller (2020) & \cite{Powell2020} & FMT & XCFC & s18np & LS220 & $18.0$ \\
   &&&& y20 & LS220 & $20.0$ \\
   \hline
   Pan et al. (2021) & \cite{Pan2021} & IDSA & TOV & NR & LS220 & $40.0$ \\
   \hline
   Powell et al. (2021) & \cite{Powell2021} & FMT & XCFC & z85\_LS220 & LS220 & $85.0$ \\
   &&&& z85\_SFHo & SFHo & $85.0$\\
   &&&& z85\_SFHx & SFHx & $85.0$\\
   &&&& z100\_SFHo & SFHo & $100.0$\\
   &&&& z100\_SFHx & SFHx & $100.0$\\
   \hline
   Mezzacappa et al (2022) & \cite{Mezzacappa2020} & MGFLD & TOV & C15-3D & LS220 & $15.0$ \\
   \hline\hline
 \end{tabular}
\end{table*}

\begin{table*}
 \centering 
 \caption{List of additional 2D CCSN simulations used in part of the analysis to calibrate the frequency range of our phenomenological waveforms. See the text for a description of the labels.  }
 \label{tab:CCSNModels2D}
 \begin{tabular}{llccllc}
  \hline\hline
   Source & $\qquad$ &  Neutrino transport & Gravity &Model & EOS & $M_{\rm ZAMS}$  \\ \hline
   Yakunin et al. (2015) & \cite{Yakunin:2015} & MGFLD & TOV & B12-WH07 & LS220 & $12.0$ \\
   &&&& B15-WH07 & LS220 & $15.0$ \\
   &&&& B15-WH07 & LS220 & $20.0$ \\
   &&&& B15-WH07 & LS220 & $25.0$ \\
   \hline
   Morozova et al. (2018) & \cite{Morozova2018} & M1 & TOV & M10\_DD2 & DD2 & $10.0$\\
   &&&& M10\_LS220 & LS220 & $10.0$\\
   &&&& M10\_LS220\_nomanybody & LS220 & $10.0$\\
   &&&& M10\_SFHo & SFHo & $10.0$\\
   &&&& M13\_SFHo & SFHo & $13.0$\\
   &&&& M13\_SFHo\_multipole & SFHo & $13.0$\\
   &&&& M19\_SFHo & SFHo & $19.0$\\
   \hline
   O'Connor \& Couch (2018) & \cite{O'Connor2018} & M1 & TOV  & mesa20\_2D & SFHo & $20.0$\\
   &&&& mesa20\_2D\_pert & SFHo & $20.0$ \\
   \hline
    Pan et al. (2018) & \cite{Pan2018} & IDSA & TOV & BHBlp & BHB$\Lambda\phi$ & $40.0$ \\
    &&&& DD2 & DD2 & $40.0$ \\
    &&&& LS220 & LS220 & $40.0$ \\
    &&&& SFHo & SFHo & $40.0$ \\
    \hline
    Eggenberger Andersen et al. (2021) & \cite{EggenbergerAndersen2021} & M1 & TOV & k200 & k200\_m0.75 & $20.0$\\
    &&&& m0.55 & k230\_m0.55 & $20.0$ \\ 
    &&&& m0.75 & k230\_m0.75 & $20.0$ \\ 
    &&&& m0.95 & k230\_m0.95 & $20.0$ \\
    &&&& k260 & k260\_m0.75 & $20.0$ \\  
   \hline
 \hline\hline
 \end{tabular}
\end{table*}
%%%%%%%%%%
\subsection{CCSN waveform catalogue}

Since we aim to design a phenomenological waveform generator for CCSN, we want to base the parameters of our waveforms (amplitude, frequency evolution, duration ...) on realistic CCSN waveforms. Therefore, we build a catalogue of selected CCSN waveforms from the literature that we use to calibrate the different parameters. The selection criteria are:

\begin{enumerate}[i)]
\item three-dimensional simulations, 
\item non-rotating progenitors, 
\item no-magnetic fields, 
\item some neutrino transport treatment (we do not consider models with light-bulb models or grey leakage schemes), 
\item progenitors with single star stellar evolution (we do not consider e.g. ultra-stripped stars result of binary interaction \cite{Powell2019}).
\end{enumerate}

Table~\ref{tab:CCSNModels} shows the selection of 33 models forming part of our catalogue. Note that we had to freeze the composition of our catalogue in 2022 since the results of this work are already being used in some other ongoing projects. Therefore, some recent simulations \cite[e.g.][]{Mezzacappa2023,Vartanyan2023} are not being used for this catalogue even if they would fit according to our selection criteria. Additionally to these models, for some of the statistics (frequency evolution), we include an additional set of 17 2D simulations (the rest of the criteria are the same). Unless explicitly said we use only the 3D simulations set, which is referred to simply as the catalogue.

The models in Table~\ref{tab:CCSNModels} use a variety of neutrino transport schemes including the Fast Multigrid Transport (FMT, \cite{Muller2015}), the Isotropic Diffusion Source Approximation (IDSA, \cite{Liebendorfer2009}), an energy-dependent first-momentum closure scheme (M1, \cite{Levermore1984}), energy-dependent two-moment equations using a variable Eddington factor technique (Ray-by-ray+,\cite{Buras2006}) and the Multigroup Flux-Limited Diffusion scheme (MGFLD, \cite{Bruenn2016}). For the treatment of gravity, different approaches are used including a pseudo-relativistic gravity treatment (TOV, \cite{Marek2006}), a conformally flat condition approximation of general relativity (XCFC, \cite{Cordero-Carrion2009}) and a fully hyperbolic formulation of general relativity (BSSN, \cite{Shibata1995,Baumgarte1998}). The equations of state used include BHB$\Lambda\phi$ \cite{Banik2014}, DD2 \cite{Fischer2014}, the k2$*$0\_m0.$*$5 series \cite{Schneider2020}, LS220 \cite{Lattimer1991}, TM1  \cite{Hempel2010}, SFHo and SFHx \cite{Steiner2013}. 

% ---- Public availability of all waveforms -----
% Kuroda'16 - not public
% Andresen 17 - https://wwwmpa.mpa-garching.mpg.de/ccsnarchive/data/Andresen2016_public/
% Kuroda '17 - not public 
% Yakunin 17  - http://eagle.phys.utk.edu/chimerasn/trac/wiki/GravitationalWave
% O'Connor & Couch - https://ttt.astro.su.se/~eoco/oconnorcouch2018b_gwdata.tar.gz
% Andresen 19 - https://wwwmpa.mpa-garching.mpg.de/ccsnarchive/data/Andresen2019
% Powel & Müller 2019 - not public
% Radice 2019 - https://www.astro.princeton.edu/~burrows/gw.3d
% Powel & Müller 2020 - not public
% Pan 2021 - not public
% Powell et al 2021 - not public
% Mezzacappa 2021 - http://eagle.phys.utk.edu/chimerasn/trac/wiki/GravitationalWave

\begin{figure}[t]
 \includegraphics[width=0.48\textwidth]{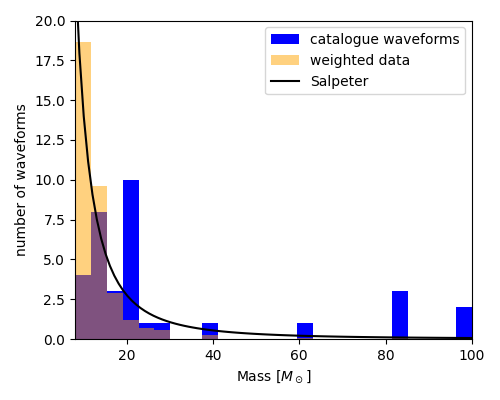}
 \caption{Distribution of waveforms in our catalogue according to the ZAMS mass (blue bars) compared to the Salpeter's initial mass function (black curve). Orange bars show the result of weighing the distributions with Salpeter's law.}
 \label{fig:ProgenitorMasses}
\end{figure}

Progenitors span zero-age main-sequence (ZAMS) masses in the range $9$ to $100$~$M_\odot$. All progenitors are generated in a solar metallicity environment except for the model from \cite{Pan2021} and those from \cite{Powell2021}, which are zero metallicity. Given that the catalogue is a selection based on what is available in the literature, the distribution of progenitor ZAMS masses does not follow what one would expect from the initial mass function (IMF) \cite{Salpeter1955}. This can be seen in Fig.~\ref{fig:ProgenitorMasses} that shows the number of waveforms in the catalogue depending on the mass compared to the Salpeter initial mass function \cite{Salpeter1955} that models the probability of having a progenitor of certain initial mass, $M_{\rm ZAMS}$ per unit mass as proportional to $M_{\rm ZAMS}^{-2.35}$ (black curve). If we use the catalogue as it is to estimate the parameters that we need for the phenomenological templates the results will be biased by the fact that there is an excess of waveforms for high-mass progenitors. To avoid this bias we introduce a weight function
\begin{equation}
    \textrm{weight} (M_{\rm ZAMS}) = C \frac{M_{\rm ZAMS}^{-2.35}}{N_{\rm waveforms} (M_{\rm ZAMS})},
    \label{eq:weight}
\end{equation}
where $N_{\rm waveforms} (M_{\rm ZAMS})$ is the number of waveforms per unit mass and $C$ is a normalization constant such that 
\begin{equation}
    \int_{8 M_\odot}^{100 M_\odot} \textrm{weight} (M_{\rm ZAMS})\, d M_{\rm ZAMS} = 1.
\end{equation}
We assign a weight to each waveform based on this function, which is used wherever we make a statistical estimation in the next subsections. As an example, we show in Fig.~\ref{fig:ProgenitorMasses} (orange bars) that the weighted mass distribution follows Salpeter's law.

The GW emission in all numerical simulations is dominated by a PNS oscillation mode that has been identified as a g-mode \cite{BMueller:2013,Cerda-Duran2013,Torres-Forne2018,Torres-Forne2019a,Torres-Forne2019b}, a f-mode \cite{Morozova2018,Sotani2019,Sotani2020} or some combination of both \cite{Vartanyan2023}. Regardless of the nature of this mode, all CCSN simulations show a high power GW emission at about a few $100$~Hz after bounce raising up to a few kHz in a timescale of $0.5-1$~s. This feature appears as a rising arch in time-frequency representations (spectrograms) and we call it {\it dominant mode} hereafter. Additionally, a sub-dominant feature present in the GW signal of many of the simulations is the imprint of the standing-shock accretion instability (SASI), appearing at low frequencies of about $100$~Hz.

For our analysis of the next subsections we use in some cases a high-pass filter at $200$~Hz that isolates the dominant mode by removing from the signal low-frequency components such as SASI modes (that are treated separately) or very low frequency ($<10$~Hz) memory effects. Similarly, to isolate the SASI modes we use a band-pass filter between $50$ and $200$~Hz, that removes both the dominant mode at higher frequencies and memory effects at lower ones. We use a second or fourth-order Butterworth \cite{Butterworth1930} filter for the dominant and SASI modes, respectively\footnote{We used the implementation in \texttt{scipy.signal.butter} of the \texttt{scipy} library \cite{2020SciPy-NMeth} freely available at \url{https://scipy.org}.}. Although the ability of the filters to isolate modes is limited and some cross-contamination is possible (especially at early times), it is sufficient for the purpose of our analysis.

%%%%%%%%%%%%%%%%%%%
\subsection{Signal duration for the dominant mode}

\begin{figure}[t]
 \includegraphics[width=0.48\textwidth]{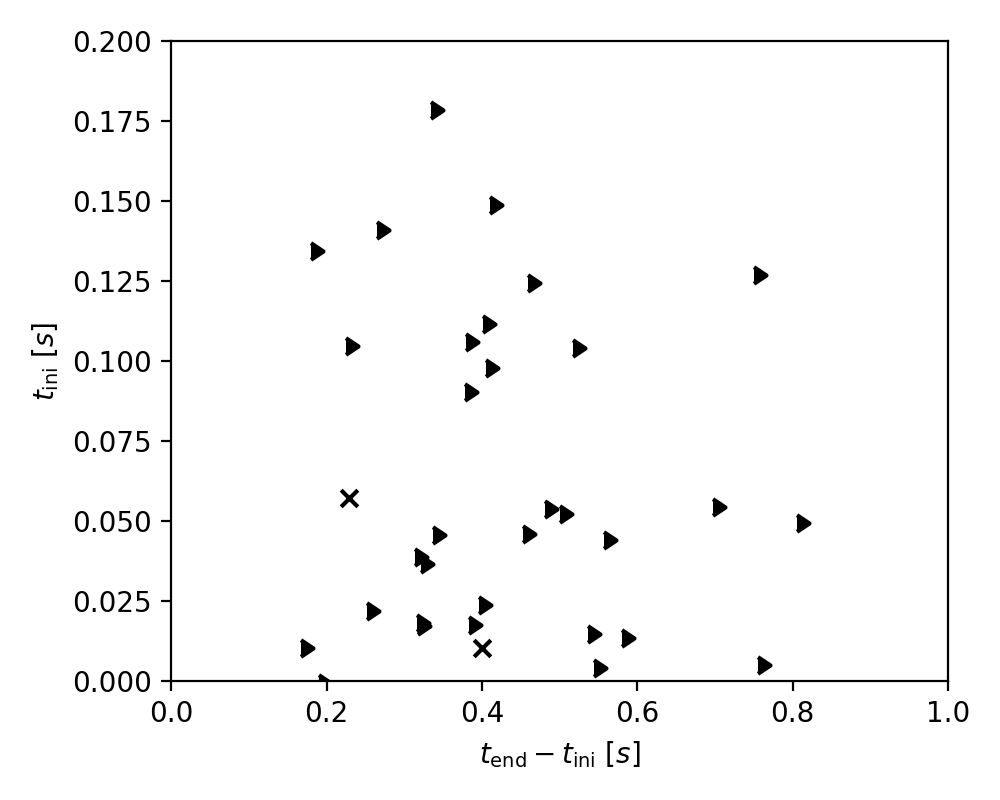}
 \caption{Post-bounce time at which the GW emission starts, $t_{\rm ini}$, as a function of the signal duration, $t_{\rm end} - t_{\rm ini}$. Triangles show lower limits for the signal duration. }
 \label{fig:StatDuration}
\end{figure}

For the case of non-rotating progenitors, the bulk of the GW emission starts shortly after the bounce and continues for about $\sim 0.2$-$1$~s up to the time of the explosion. After the explosion, the GW amplitude decreases significantly although the dominant mode can still emit for periods of time as long as $6$~s \cite{Vartanyan2023}. For non-exploding models, the GW emission keeps at high amplitude levels \cite{Cerda-Duran2013,Vartanyan2023} of amplitude for longer timescales stopping abruptly at the time of black hole formation \cite{Cerda-Duran2013}. For our model, we focus on the initial dominant part of the emission.

Hereafter, we use $t$ as a measure of the post-bounce time, such that $t=0$ corresponds to the bounce time. Note that, since we are considering non-rotating progenitors, there is no rotationally-induced deformation of the collapsing core and hence there is no bounce signal associated with this time. The GW emission starts at a later time $t_{\rm ini}\ge 0$ when the different instabilities break the spherical symmetry of the system. The delay between the start of the GW emission and the bounce can be close to zero, but can also be as long as $0.25$~s \citep[see e.g.][]{Mueller:2012}.
We consider that the GW emission stops when its amplitude decreases significantly with respect to its maximum value. We define the time at which this occurs as $t_{\rm end}$.  The duration of the signal is therefore $t_{\rm end} - t_{\rm ini}$

It is complicated to make accurate statistics of the duration of the signal because many numerical simulations stop at fixed times and GW emission may not have finished at that time. The longest simulation of our catalogue is the model y20 of \cite{Powell2020} with a duration of $1.2$~s. However, the emission for this particular model decreases significantly after $0.6$~s. On the other end, some of the zero metallicity progenitors of \cite{Powell2021} have very short-lived GW emission of $\sim 0.2$~s and low-mass progenitors typically have durations of $0.2-0.3$~s too (e.g. the $9$~$M_\odot$ model in \cite{O'Connor2018}).

To estimate the duration of the dominant mode emission in the waveforms of the catalogue we define the integrated signal energy as
\begin{equation}
    E_{\rm signal} (t) = \int_0^t |h(t')|^2 dt'.
\end{equation}
We define the measured values of $t_{\rm ini}$ and $t_{\rm end}$ as the times in which $E_{\rm signal}$ corresponds to $5\%$ and $95\%$ of the total signal energy, respectively. To ensure that the measured duration corresponds to the dominant mode emission (see discussion below) we apply a high-pass filter to the waveform at $200$~Hz. In most of the cases, the simulation was stopped when the GW emission was still significant. To detect such cases we compute the signal power for $t>t_{\rm end}$ as the signal energy in this interval over the time interval. For cases in which the signal power is larger than $0.1\,$ we consider that a significant GW amount of signal energy is missing. In those cases, we take the end time of the simulation as a lower limit to $t_{\rm end}$. Fig.~\ref{fig:StatDuration} shows the value of the delay $t_{\rm ini}$ as a function of the duration of the waveforms (mostly lower limits). The obtained values match what is described qualitatively above.

%%%%%%%%%%%%%%%%%%%
\subsection{Frequency evolution of the dominant mode}
\label{sec:FreqEvol}

As the background slowly changes in time due to accretion and PNS cooling, the value of $\omega_0$ for the different
eigenmodes will change with time. To have a more direct comparison with results from simulations we use the frequency, $f=\omega_0/(2\pi)$, for most of the discussion hereafter. This subsection aims to find a function $f (t)$ depending on a few parameters that can describe the diversity mode-frequency evolutions observed in simulations. We focus first on the dominant mode.

To describe the frequency evolution we use a similar approach to \cite{Lopez2021}, defining $f(t)$ as an interpolation to a series of $n_p$ discrete points $\{(t_i,f_i)\}$,  $i = 1, ..., n_p$.
$f(t)$ is constructed following the next procedure
\begin{enumerate}
\item In the interval $[t_0,t_{n_p}]$ we interpolate the function using a piecewise cubic splines interpolation (Steffen method, \citep{Steffen1990}) that guarantees monotonicity of the interpolating function. For the case $n_p=2$ we use linear interpolation.
\item Outside $[t_0,t_{n_p}]$  but inside $[t_{\rm ini},t_{\rm end}]$  we use linear extrapolation.
\item Outside the interval $[t_{\rm ini}, t_{\rm end}]$ we set a constant value of $f$ matched continuously to the previously generated function. The waveform is not excited in this region (the strain is essentially zero) but setting this value allows for a smooth transition when generating the waveforms. 
\end{enumerate}
For all the interpolations we use the implementation in the \texttt{GNU} scientific library, \texttt{GSL}\footnote{\url{http://www.gnu.org/software/gsl}} \cite{gough2009gnu}.

Note that the times $t_i$ used to generate $f(t)$ do not have to be actual times in which the signal is excited, but just serve to generate the function. That means that $t_i$ does not have necessarily to fall in the interval  $[t_{\rm ini}, t_{\rm end}]$.

\begin{figure}[t]
 \includegraphics[width=0.49\textwidth]{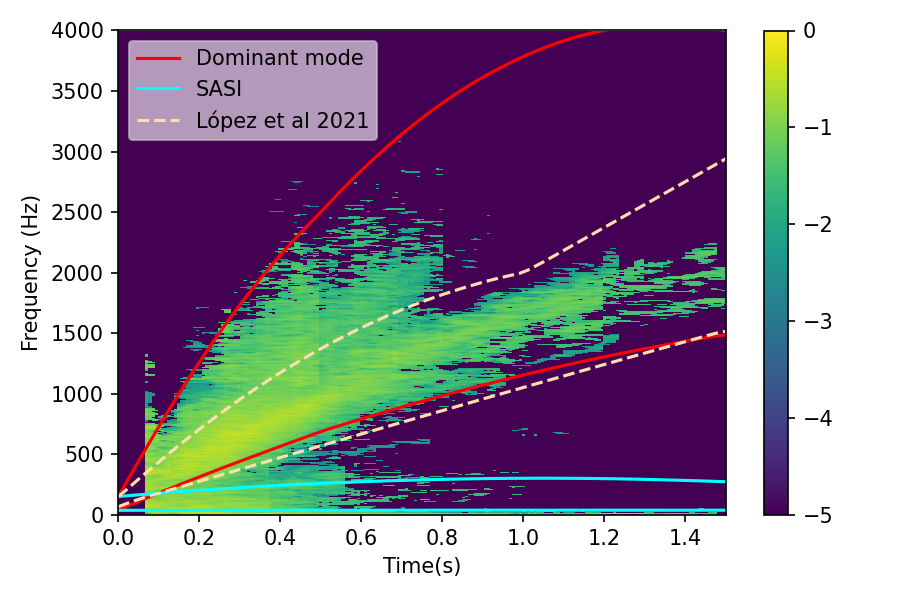}
 \caption{Averaged spectrum of a set of 2D and 3D simulations weighed by the Salpeter's IMF. Color-coded the logarithm of the spectral density normalized to the maximum frequency at each time. 
Red and cyan curves indicate the lower and upper limits of the regions covered by our phenomenological templates for the dominant mode and SASI, respectively. For comparison, we plot the region covered by \cite{Lopez2021} (light orange curves). }
 \label{fig:StatSpectrogram}
\end{figure}

To determine the possible range of values for $f(t)$ we compute spectrograms for all the waveforms in the catalogues described in Section~\ref{sec:PNSoscillations} including both 2D and 3D simulations, for additional statistics. We have removed the waveforms of \cite{Andresen:2017,Andresen2019} that are sampled at $1$~kHz which results in important aliasing artefacts in the spectrograms. For many of the waveforms, the data provided allows us to compute the waveform at any observing angle. In those cases, we generated $200$ waveforms for each simulation and averaged results over all angles. As a summary of all simulations, Fig.~\ref{fig:StatSpectrogram} shows the average spectrum of all simulations at all angles weighted by Salpeter's IMF. Since we are only interested in the frequency evolution, and not the strain itself, the final result is normalized to the maximum amplitude at each time. We can see that the dominant mode starts after bounce at low frequencies of $100-500$~Hz, raising up to $500-2500$~Hz at $0.6$~s. At later times the frequency keeps rising. However, given the small number of simulations reaching such large simulation times, there is not much statistics on the possible values. One could imagine that the simulations with frequencies of $2000$~Hz at $0.6$~s would keep raising their frequency probably up to $3$ or $4$~kHz. The light orange curves mark the limit of possible $f(t)$ used in \cite{Lopez2021}, which can describe the bulk of CCSN simulations. Red curves mark the limit of possible values of $f(t)$ used in this work (see description on Section~\ref{sec:ccphen}) that allows for more extreme possibilities observed in some numerical simulations. 

In our work, these extreme models with higher frequencies correspond to three of the four models of \cite{Pan2018} (BBH$\Lambda\phi$, LS220 and SFHo) and the NR model in \cite{Pan2021}. All of these models use the s40 progenitor from \cite{Woosley2007}, which has a massive and compact iron core, combined with relatively soft EOS ($M_{\rm max}\sim 2.1 M_{\odot}$ in all cases). This  
combination produces black holes in very short timescales ($\sim0.6$~s for LS220 according to \citep{O'Connor2018}). This means that the compactness of these models evolves very quickly, leaving an imprint in the spectrogram with a large slope in the time-frequency evolution. This same progenitor with a stiffer EOS (DD2, $M_{\rm max}=2.42 M_{\odot}$) does not produce such a fast rise in frequency, and remains within the frequencies considered in \cite{Lopez2021}.

It is also worth noticing that the mode frequencies obtained in numerical simulations may depend on the gravity treatment used. \cite{Westernacher-Schneider2019} has noted that the use of a pseudo-Newtonian potential (marked as TOV in Tables~\ref{tab:CCSNModels} and \ref{tab:CCSNModels2D}) can produce differences of $\sim10\%$ in the frequencies with respect to general relativistic simulations (those marked as XCFC or BSSN in the tables). These differences are not very relevant for the purpose of this work, which only uses them to estimate the possible range of values of the frequency, but it may have a large impact if the frequencies are used for asteroseismology.

%%%%%%%%%%%%%%%%%%%
\subsection{Q-factor of the dominant mode}

The oscillations in the PNS are in general damped quickly due to non-linear coupling and the lack of adiabaticity of
the oscillations (cooling and heating processes), leading to a negative value of oscillation frequency $\sigma$ (see Section \ref{sec:oscillations}). The GW emission itself
is a slow process acting in timescales of $1-10^5$~s \citep{Torres-Forne2019a,Sotani2020}, so it can be neglected here. 
All this complicated physics is encompassed in a single $Q$ factor. For values of $Q<1/2$ oscillations would be overdamped, and GW emission would be strongly suppressed. Given that this is not observed in numerical simulations we consider here only values of $Q>1/2$. The presence of a damping term broadens the spectrum of the real oscillations of the system, which becomes narrower as $Q$ increases. The full width at half maximum of the spectrum can be estimated as $\Delta f = f / Q$. Therefore, estimating how broad is the spectrum of the modes in the GW signal from numerical simulations it is possible to estimate the value of Q. Typical values of $\Delta f / f$ in numerical simulations are in the range $0.1$-$1$, which implies 
values of $Q \in [1,10]$. The value of $\Delta f / f$ does not appear to change significantly in time. Therefore, for simplicity, we consider that the Q-factor is constant in time. 

%%%%%%%%%%%%%%%%%%%
\subsection{Strain amplitude of the dominant mode}
\label{sec:par:amplitude}

In general, the amplitude of the dominant mode can vary with time. We quantify the amplitude of the signal by means of the rms angular-averaged value, $\bar h_{\rm rms}$, as defined as in Eq.~\eqref{eq:hrms}. As defined there, it is a global quantity of the waveform that gives an idea of the average strain value. However, it can also be computed in small intervals $[t-\Delta t, t+\Delta t]$ to define an instantaneous $\bar h_{\rm rms} (t)$. For this work, we consider constant $\bar h_{\rm rms} (t)$.

\begin{figure}[t]
 \includegraphics[width=0.48\textwidth]{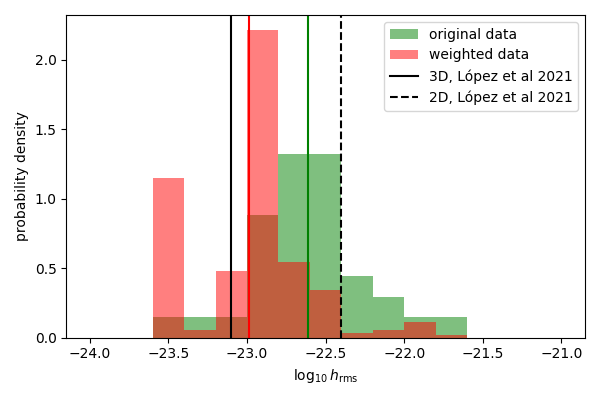}
 \caption{Probability density of the logarithm of the rms angular-averaged strain for the selected catalogue of 3D simulations in Table~\ref{tab:CCSNModels} (green bars) and for the same data but weighted by Salpeter's IMF. Vertical green and red lines correspond to the median of each of the datasets, respectively. For comparison, we show the mean values for 3D and 2D data in \cite{Lopez2021}. Given the counts of the $i$th bin $c_i$ and its width $b_i$, we define the probability density  as $c_i/(\sum_i^N c_i \times b_i)$, where $N$ is the total number of bins of the histogram.}
 \label{fig:Stathrms}
\end{figure}

For the case of constant amplitude, and similarly to \cite{Lopez2021}, we face the problem of what is the strain amplitude of a typical CCSN observed at a certain distance $D$. We use the catalogue of 3D simulations of Table~\ref{tab:CCSNModels} to calibrate the distance-strain relation. Green bars in Fig.~\ref{fig:Stathrms} show the distribution numerical simulations according to their value of $\log_{10} \bar h_{\rm rms}$ for sources at $D=10$~kpc. $\bar h_{\rm rms }$ has been computed to the signal with the $200$~Hz high-pass filter to include only the contribution of the dominant mode. As we noted above, this distribution is probably biased by the selection of progenitor masses in the simulations. To try to eliminate this bias we have weighted the data with Salpeter's IMF as described above (see Eq.~\eqref{eq:weight}). The resulting distribution (red bars)  has a median value and standard deviation of
\begin{equation}
    \log_{10} \bar h_{\rm rms}=-23.0 \pm 0.4
    \label{eq:loghrms}
\end{equation}
at $10$~kpc. The values obtained here are similar to those in \cite{Lopez2021}. The reason is twofold. On the one hand, the simulations used in our work use in general more sophisticated neutrino transport than the ones in \cite{Lopez2021} and in many cases better numerical resolution. This gives in general an increase in the rms strain amplitude. On the other hand, the weighting by Salpeter's IMF decreases the median value of the distribution to a level similar to \cite{Lopez2021}. It is important to notice that there are important differences in the rms strain between 2D and 3D simulations \cite{Lopez2021}, being the former systematically larger. So it is important only to use 3D simulations in the calibration.

%%%%%%%%%%%%%%%%%%%
\subsection{Forcing for the dominant mode}

As we discussed above, the spectral width of the dominant mode (related to the $Q$ factor) indicates that modes in the PNS are quickly damped ($Q=1-10$ implies damping times of $2Q/\omega\sim 0.3-3$~ms at $1$~kHz). Therefore, the dominant mode feature observed in spectrograms for hundreds of milliseconds to seconds implies that this mode is being continuously excited. \cite{Murphy:2009, Mezzacappa2023} suggested that the excitation of PNS modes was due to impulsive hits of matter that fall asymmetrically onto the PNS, either due to the presence of post-shock convection or to the SASI. Alternatively, \cite{Mezzacappa2023} has suggested the origin of the excitation of the modes convection in the PNS interior and overshooting. Regardless of the physical origin of the excitation of the modes we can prescribe a forcing term in Eq.~\eqref{eq:harmonic} that mimics this excitation.  

To compute the forcing energy $W_{lm}(t)$ needed in Eq.~\eqref{eq:harmonic} we first compute the forcing power $P_{lm}(t)$ and then integrate numerically Eq.~\eqref{eq:power}. The amplitude of the forcing sets the amplitude of the waveform therefore one has to be careful with the normalization. A global normalization to the waveform amplitude is discussed in Section~\ref{sec:par:amplitude}; here we just discuss the dependency of the amplitude on the frequency. With an equation analogous to Eq.~\eqref{eq:harmonic}, \cite{Lopez2021} showed that if one applies a random forcing with a constant average value the resulting strain amplitude depends on the frequency. To obtain a statistically constant amplitude regardless of the frequency they had to scale the forcing with $f(t)^{-0.45}$. One critical difference between the work of \cite{Lopez2021} and ours is that they integrate directly the strain at the detector, while we are obtaining $I'_{lm}$ so the strain is a second-time derivative of this quantity. To have the same behaviour for our strain we have to scale the power forcing by $f(t)^{1-0.45}$, so that the forcing energy scales as $f(t)^{2-0.45}$ and hence the second time derivative of the strain scales as $f(t)^{-0.45}$, as in the case of \cite{Lopez2021}. The derivation of these exponents is purely phenomenological based on the behavior of the code. 

Our forcing power has the form 
\begin{equation}
P_{lm}(t) \propto \textrm{random}(t) \cdot f(t)^{1-0.45} \cdot \bar h_{\rm rms} (t),
\end{equation}
where $\textrm{random} (t)$ is Gaussian white noise generated using \texttt{GSL}\footnote{We use the default random number generator in \texttt{GSL},  MT19937 \cite{MT19937}, with a repetition period of about $10^{6000}$, more than sufficient for our application. 
The impact of the random number generator in the CPU time has been tested to be $<1\%$. }. If the function $\bar h_{\rm rms} (t)$ is constant, then the resulting waveform resulting from the use of $P_{lm} (t)$ will have a constant amplitude (in a statistical sense). In case $\bar h_{\rm rms}(t)$ is not constant, the rms value of the resulting waveform will follow this function. 

As we show in the next sections this approach reproduces qualitatively the dominant mode observed in the GW signal of simulations and is able to set the evolution of the rms values of the signal at will. 

Alternatively, to the forcing power described here we have tested two alternatives: i) We have tested a forcing term similar to \cite{Lopez2021}. This contains a free parameter that, if tuned appropriately, gives results very similar to the results given here. However, this extra parameter does not add any additional advantage. Overall, our method is easier to implement and does not need this extra parameter to tune up. ii) The second possibility that we tested we a random Gaussian noise colored by a normal distribution of a certain mean frequency and width. This is motivated by the spectrum of PNS convection that was observed by \cite{Raynaud2022}. Although the approach may seem to have a better physical motivation, the results are actually quite poor. Eq.~\ref{eq:harmonic} acts as a resonator, so if the forcing is only applied in part of the frequency spectrum, the dominant mode is only excited if the frequency $f(t)$ lays within the excitation interval. The result is that the dominant mode can only be excited for its full duration if the width of the Gaussian is sufficiently large. However, this case is effectively almost the same as applying white noise. So again, the extra complication and parameters do not add any extra information. Therefore, these two alternatives have been abandoned and are discussed here just as things not to do. A lesson that we learn is that, if we want to excite the same dominant mode for timescales of hundreds of milliseconds to seconds, the forcing spectrum has to be spectrally very broad.

%%%%%%%%%%%%%%%%%%%
\subsection{Standing shock accretion instability (SASI)}
\label{sec:SASI}

In case the SASI is active, it produces a GW signature in the spectrograms similar in many senses to the ones of the dominant mode. To model SASI modes we use the same formalism as for the dominant mode but adapt the parameters to the properties of SASI. Since the SASI can only be active before the onset of the explosion, the range of possible durations is the same as for the dominant mode. Similarly happens with $Q$, with typical values also in the range $1-10$. We use the same forcing model used for the dominant mode. 

The frequency evolution is quite different, with a typical value of $50-150$~Hz after bounce raising linearly up to $50-300$~Hz. The rms strain is typically a fraction of the one of the dominant mode. In extreme cases, the amplitude can be comparable to the dominant mode while in others is very weak or non-existent.

%% file: ccphen.tex
%%%%%%%%%%%%%%%%%%%
\section{Phenomenological templates}
\label{sec:ccphen}

\begin{figure*}[t]
    \includegraphics[width=0.32\textwidth]{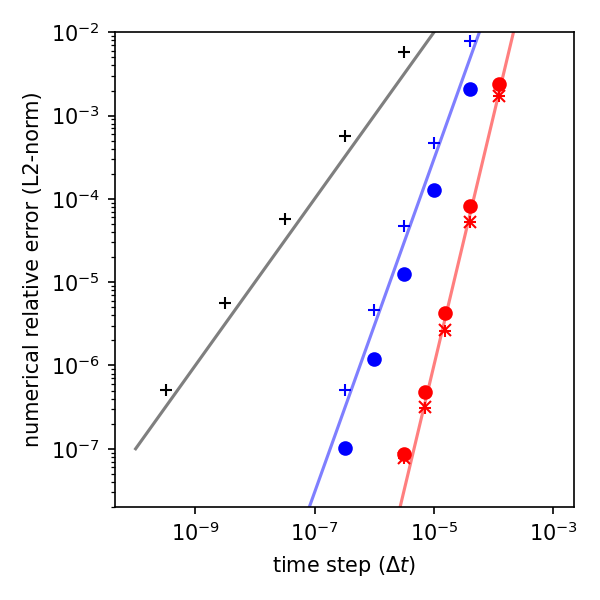}
    \includegraphics[width=0.32\textwidth]{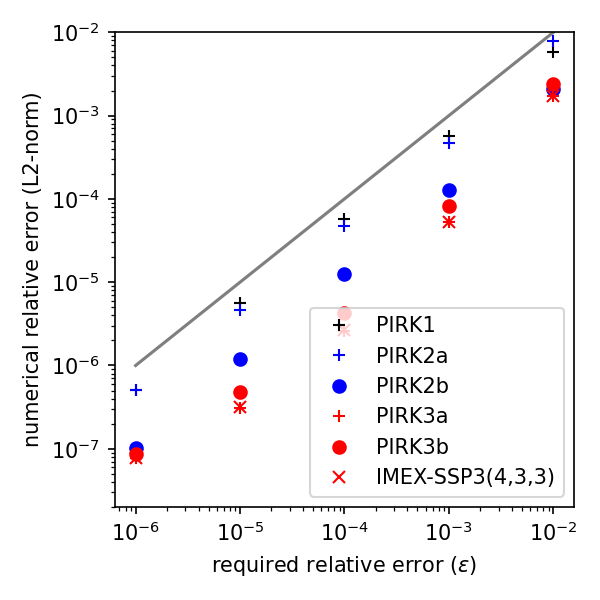}
    \includegraphics[width=0.32\textwidth]{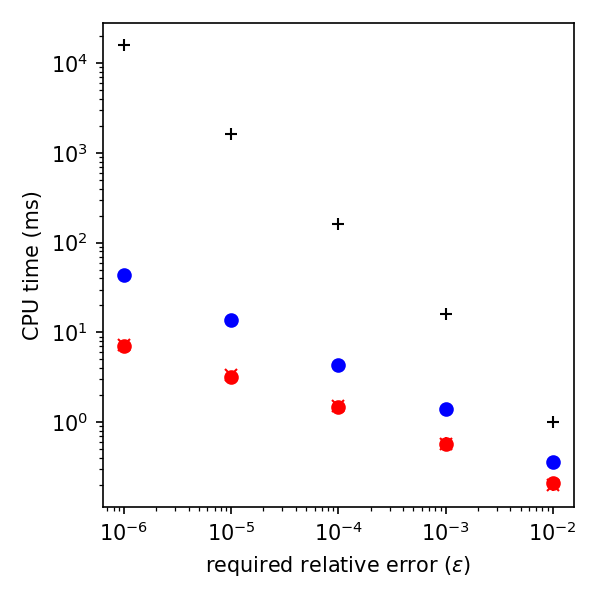}
 \caption{Performance of the different time integration methods: numerical relative error as a function of the time step (left panel) and required relative error (central panel). Black, blue and red lines in the left panel have slopes of $1$, $2$ and $3$ respectively, corresponding to the order of the methods used. The black line in the central panel represents the numerical relative error equal to the relative error. The left panel shows the dependence of the CPU time with the required relative error.}
 \label{fig:integrator}
\end{figure*}

\begin{figure*}[t]
    %\centering
    \raggedright
    \hspace{0.0035\textwidth}
    \includegraphics[width=0.4\textwidth]{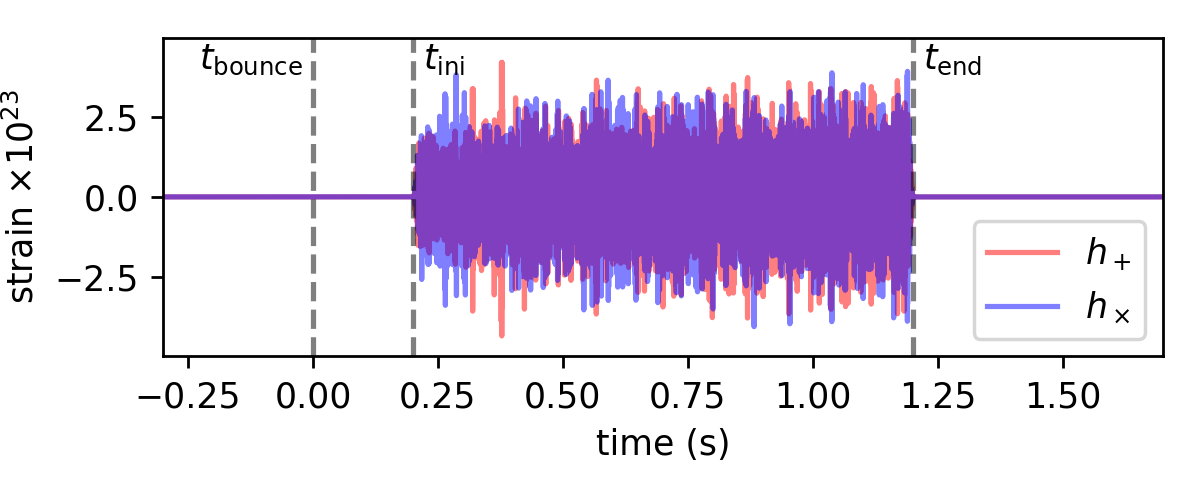}
    \hspace{0.075\textwidth}
    \includegraphics[width=0.405\textwidth]{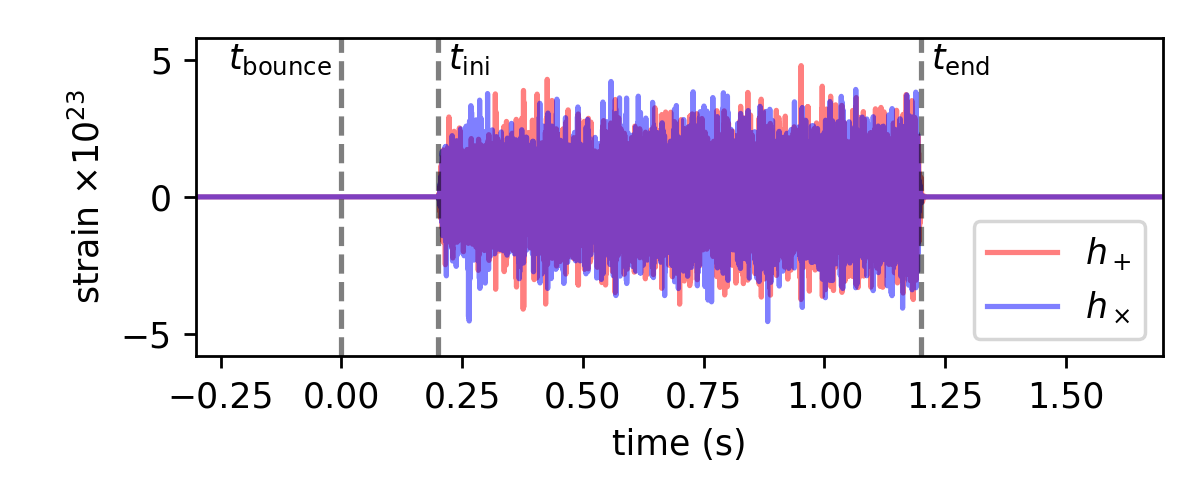}\\
    \includegraphics[width=0.49\textwidth]{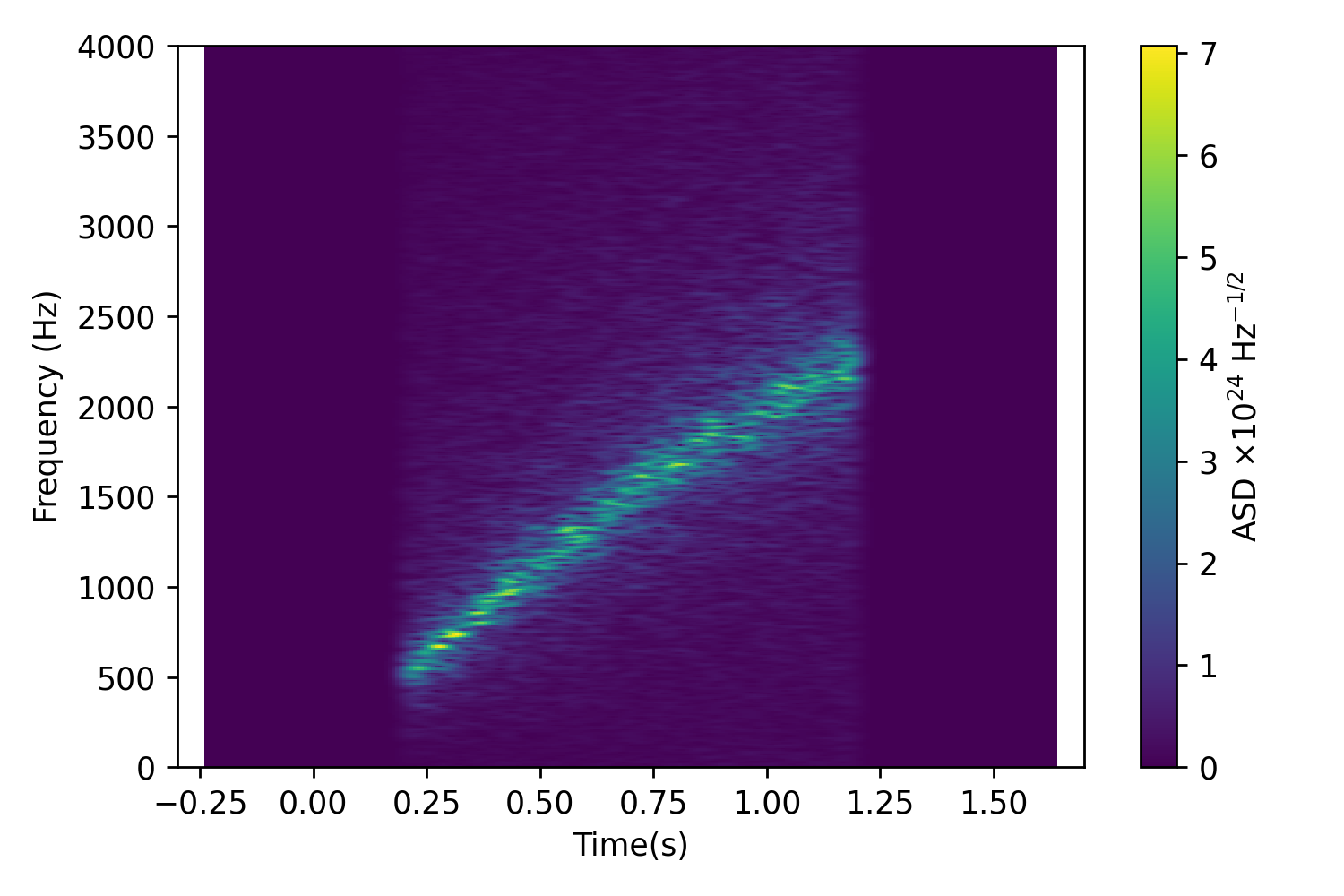}
    \includegraphics[width=0.49\textwidth]{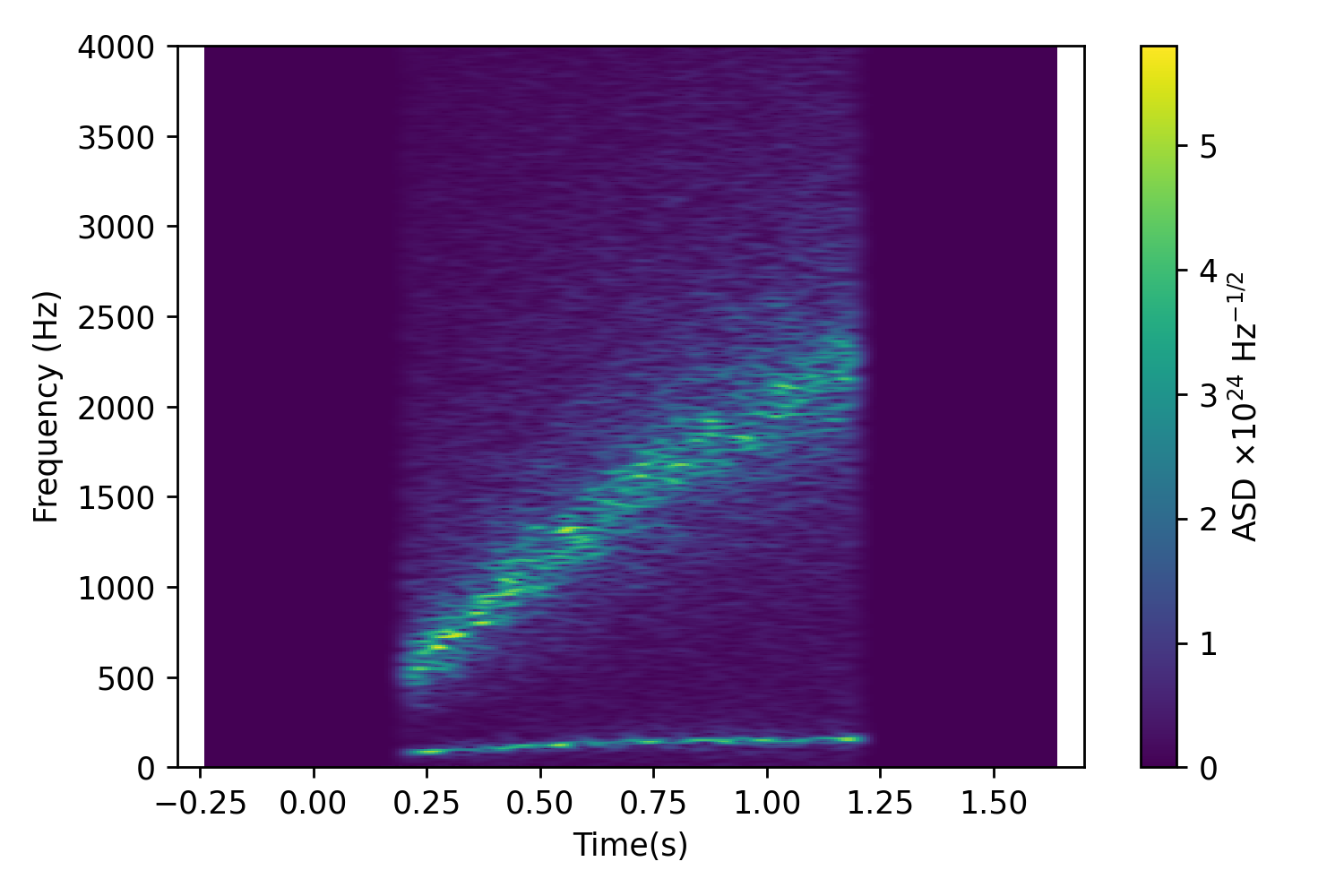}\\
    \includegraphics[width=0.49\textwidth]{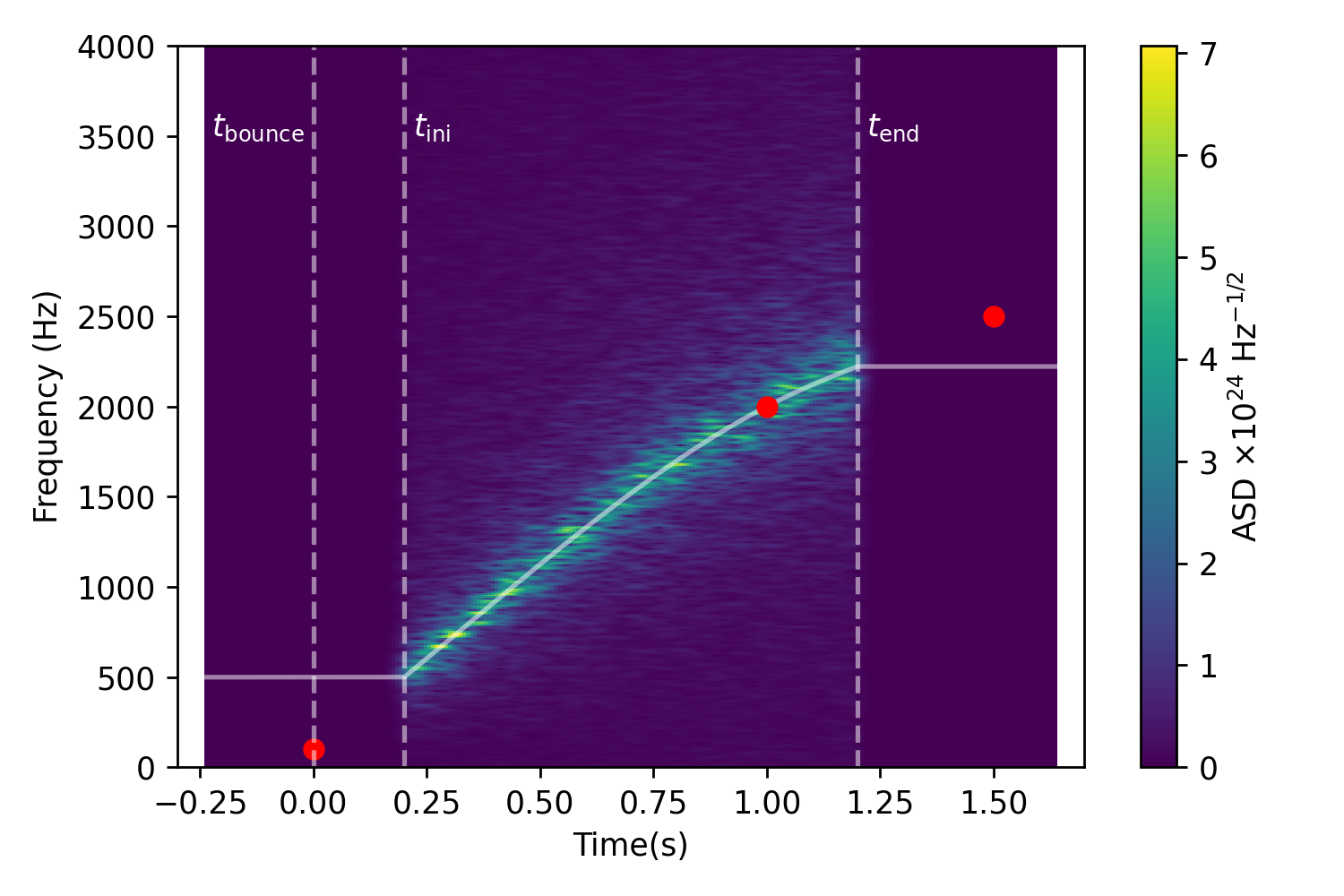}
    \includegraphics[width=0.49\textwidth]{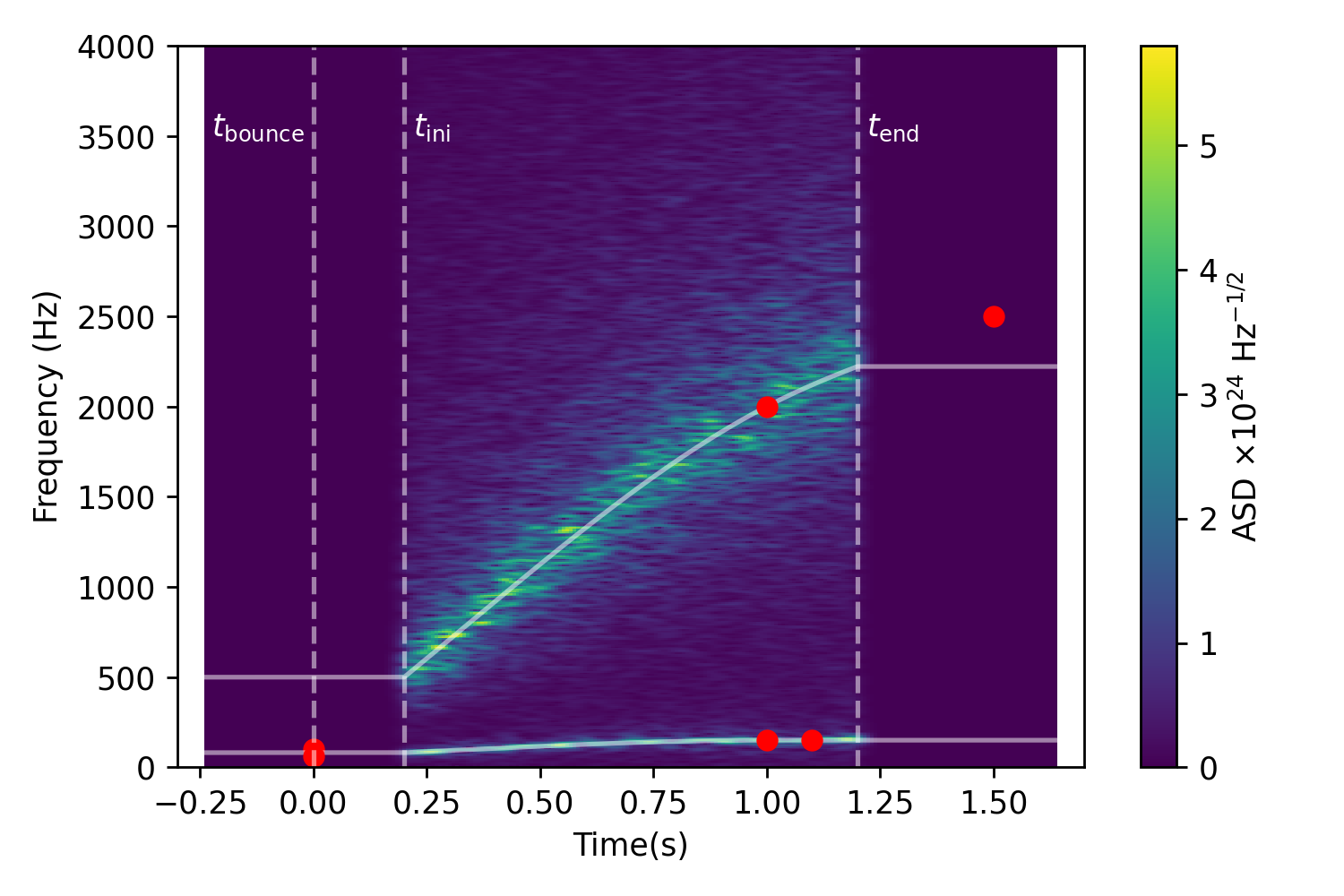}
    \caption{Examples phenomenological waveforms containing only the dominant mode (left panels) and also SASI (right panels). Upper panels show the strain. The middle and lower panels show the corresponding spectrograms. Colour-coded, the amplitude spectral density (ASD). Lower panels show the same spectrograms as the middle panels but with the prescribed frequency evolution for each component on top and the location of bounce, start and end of the waveform. Red dots correspond to the interpolation points $(t_i,f_i)$.}
    \label{fig:waveform_examples}
\end{figure*}

The phenomenological template generator described in the previous sections has been coded in \texttt{C} for fast and efficient computation, with a \texttt{python} interface for ease of use and integration in other codes and pipelines. The code, \texttt{ccphen (v4)}\footnote{Previous versions of \texttt{ccphen} correspond to the ones used in \cite{Astone2018} (\texttt{v2}) and \cite{Lopez2021} (\texttt{V3}). \texttt{v1} is a test code that was never used in any publication.}, is publicly available at \url{https://www.uv.es/cerdupa/codes/ccphen/}. 

The parameters needed by \texttt{ccphen} to generate the waveform and the components present in it (dominant mode and SASI ) can be classified as:
\begin{itemize}
\item {\it Template parameters}: sampling frequency, $f_s$, and number of samples, $N$, which fixes the total duration of the data segment to $N/f_s$.
\item {\it Source extrinsic parameters}: distance to source, $D$, and observation angles: $\Theta$ (inclination) and $\Phi$ (polarization angle).
 \item {\it Source intrinsic parameters}: beginning time and end times, $t_{\rm ini}$ and $t_{\rm end}$, $Q$ factor and the parameters controlling the frequency evolution (see Sect.~\ref{sec:FreqEvol}), which include the number of segments $n_p$, the interpolation points $\{t_i\}$, the corresponding frequencies $\{f_i\}$ and waveform amplitudes $\{h_i\}$. For this work we use $n_p=3$, $\{t_i\}=\{0, 1, 1.5\}$~s   and $\{h_i\}=\{1,1,1\}$,  but more complex waveforms could be generated. This choice of $\{h_i\}$ produces a waveform with constant amplitude over time. Additionally, it is needed to provide the $\log_{10} h_{\rm rms}$ value of the signal at $10$~kpc and its standard deviation, as described in Sections~\ref{sec:par:amplitude} (for the dominant mode) and \ref{sec:SASI} (for SASI).

 \item {\it Random number generator seed:} Additionally, it is needed an integer that serves as a seed for the random number generator used in the time integration as well as for the generation of the amplitude. This seed allows the reproduction of the waveforms in different calls of the function. To generate different realizations of the same parameters one just has to change this seed. 
\item {\it Numerical integration parameters:} the time integration method and the relative error required, $\epsilon$, as described in the next section. It is not necessary to set up these parameters in general since the default values that we provide give sufficiently accurate results for most applications (see discussion below). 
\end{itemize}

The output of the routine contains $h(t)$ for the corresponding component. Different components can be added linearly to create more complex waveforms, e.g. to create a waveform with a dominant mode and SASI. By default, the waveform is returned centred in the data segment, but if a time array is provided (of the same length) it will use that instead. 

We describe next the practical implementation of the code and the test that we have performed for validation and to assess its performance and accuracy.

%%%%%%%%%%%%%%%%%%
\subsection{Time integration method and performance}

The calculation of the waveforms requires the numerical time integration of Eq.~\eqref{eq:harmonic}. We have implemented several time integration methods: a first-order symplectic-Euler method \cite{DeVogelaere:1956}, second and third partially implicit Runge-Kutta methods \citep{Cordero-Carrion2012} and the IMEX-SSP3(4,3,3) method \citep{Pareschi2010}. All these methods have shown good stability properties when applied to oscillatory problems (see \cite{Cordero-Carrion2012} for a comparison). To ensure high accuracy we do not integrate with the time step corresponding to the desired sampling frequency ($f_{s}$) but with a smaller variable time step
\begin{equation}
\Delta t = \frac{\epsilon^{1/o}}{\omega},
\end{equation}
where $\epsilon$ is the required error and $o$ is the order of the time-integration method. To test the accuracy of the integrator we compared the waveforms generated with different values of $\epsilon$ with a reference high accuracy waveform that was computed using the first-order symplectic-Euler method and $\epsilon=10^{-7}$. This test is performed for a $1$~s long signal sampled at $8192$~Hz. Fig.~\ref{fig:integrator} shows the test we have performed for each of the time integration methods. The numerical relative error plotted in the left and central panels corresponds to the L-2 norm of the difference between the numerical solution and the reference solution. The left panel shows that the error decreases with decreasing time steps as expected for the order of each method. The central panel shows that the actual numerical error is always smaller than the required relative error, $\epsilon$, so the latter is a good upper bound. In general, given a required error, the numerical error decreases with increasing order of the method, being the smallest for the PIRK3a method of \cite{Cordero-Carrion2012} and for  IMEX-SP3 (4,3,3). Regarding CPU time (right panel), the higher the order of the method, the lower the CPU time, behaving all third-order methods in a similar way. For the standard setup, we use the PIRK3a method and $\epsilon=10^{-3}$ which gives a good trade-off between performance and accuracy.  All CPU times are computed in a desktop computer with an Apple M2 Pro processor. When considering the full waveform generator typical CPU times are in the range of $4-15$~ms for second of signal generated for sampling rates in the range $4-16$~kHz. The amount of memory used is of the order of MB, negligible when compared to full numerical simulations of CCSNe.

\begin{table*}
    \caption{Parameter ranges used to generate waveforms. The cases considered correspond to the dominant mode and the SASI mode for standard neutrino-driven supernovae, and the dominant mode for short neutrino-driven supernovae case. The parameter range of the extrinsic parameters is also provided, which is common to all cases.}
    \begin{tabular}{l@{\hskip 0.5cm}l@{\hskip 0.5cm}l@{\hskip 0.5cm}l}
        \hline \hline
         Case & Parameter & Range  & Restrictions \\  
         \hline Dominant mode (standard) &&\\ 
         &$t_{\rm ini}$& $0$ - $0.25$~s & \\
         &$t_{\rm end}$& $0.2$ - $1.5$~s & $t_{\rm end} - t_{\rm ini} > 0.4$~s \\
         &$Q$ & $1$ - $10$ & \\
         &$n_p$ &3&\\
         &$\{ t_i\}$ &$\{0,0.5,1.5\}$& \\
         &$f_1$ &$50$ - $150$~Hz & $f_1<f_2$ \\
         &$f_2$ &$700$ - $2500$~Hz & $f_2<f_3$  \\
         &$f_3$ &$1500$ - $4000$~Hz & $(f_2-f_1)/(t_2-t_1) > (f_3-f_2)/(t_3-t_2)$ \\
         \hline SASI (standard) &&&\\
         &$t_{\rm ini}$& - & Same as dominant mode\\
         &$t_{\rm end}$&- & Same as dominant mode \\
         &$Q$ & $1$ - $10$ & \\
         &$n_p$ &3&\\
         &$\{ t_i\}$ &$\{0,1,1.1\}$& \\
         &$f_1$ &$50$ - $150$~Hz & $f_1<f_2$ \\
         &$f_2$ &$50$ - $300$~Hz &   \\
         &$f_3$ & - & $f_3=f_2$ \\
         \hline Dominant mode (short) &&\\ 
         &$t_{\rm ini}$& $0$~s & \\
         &$t_{\rm end}$& $0.2$ - $0.4$~s &  \\
         &$Q$ & $1$ - $10$ & \\
         &$n_p$ &3&\\
         &$\{ t_i\}$ &$\{0,0.5,1.5\}$& \\
         &$f_1$ &$50$ - $150$~Hz & $f_1<f_2$ \\
         &$f_2$ &$700$ - $2500$~Hz & $f_2<f_3$  \\
         &$f_3$ &$1500$ - $4000$~Hz & $(f_2-f_1)/(t_2-t_1) > (f_3-f_2)/(t_3-t_2)$ \\
         \hline Extrinsic parameters &&\\
         & $\cos \Theta$ & $-1$\quad -\quad $1$ & \\
         & $\Phi$ & $0$\quad -\quad $2\pi$ \\
         \hline\hline 
    \end{tabular}
    \label{tab:Parameters}
\end{table*}

%%%%%%%%%%%%%%%%%%
\subsection{Waveform examples}

Fig.~\ref{fig:waveform_examples} shows two examples of phenomenological waveforms, one including only the dominant mode (left panels) and a second including an additional component for the SASI (right panels). In both cases, we impose a delay in the start of the GW signal ($t_{\rm ini}=0.2$~s and a total duration of $1$~s ($t_{\rm end}=1.2$~s). The lower panels show the frequency evolution, $f(t)$, used to generate the waveform. The red dots represent the pairs $(t_i,f_i)$ used to generate $f(t)$. Note that outside the range of duration of the waveform $f(t)$ is set to constant so the the actual frequency evolution does not go through all the points, although the extrapolation of the curved section does. In the case of SASI, there are two sets of $f(t)$ and of points, one for each of the waveform components. The track left in the spectrogram by the waveform follows precisely $f(t)$ and has the duration required by the input parameters of the example (plotted as vertical dashed lines).

%%%%%%%%%%%%%%%%%%
\subsection{Random waveform generator}
\label{sec:randomWfmGen}

We also provide scripts that work as random parameter generators producing automatically waveforms consistent with the parameter ranges discussed in Sect.~\ref{sec:Parameters}. With those, it is very easy to set up large databases of waveforms representative of what is expected for neutrino-driven explosions.

\begin{figure}
    \centering
    \includegraphics[width=0.99\linewidth]{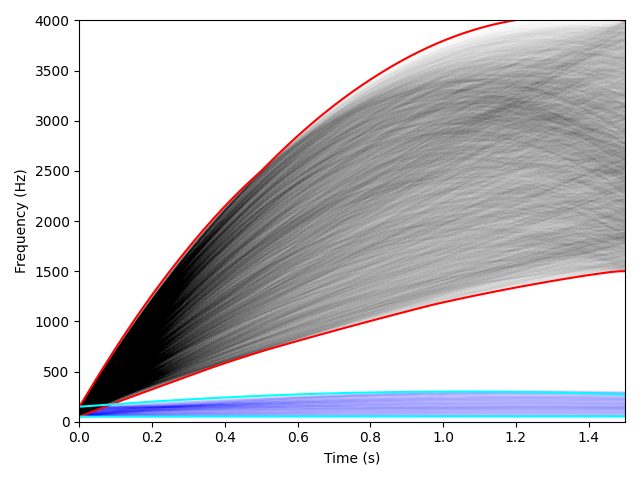}
    \caption{Frequency tracks generated using the random waveform generator for the dominant component ($10^4$ realizations, black curves) and the SASI component ($10^3$ realizations, blue curves). Red and cyan curves indicate the lower and upper limits of the regions covered by our phenomenological templates for the dominant mode and SASI, respectively.   }
    \label{fig:freq_tracks}
\end{figure}

The parameter ranges used for these generators are provided in Table~\ref{tab:Parameters}. 
We use a uniform probability distribution to generate each of the parameters within the range described in the table. Note that some of the parameters contain restrictions (e.g. the frequency). If one simply generates triads of frequencies $(f_1,f_2,f_3)$ and then discards the combinations not fulfilling the restrictions the resulting distribution of frequencies is not uniform and is strongly biased towards high frequencies. Instead of that, we generate $f_2$ with a uniform distribution, and then a duplet $(f_1,f_3)$ to which we apply the restrictions. The resulting distribution for $f_1$ and $f_3$ is much more uniform using this procedure. Fig.~\ref{fig:freq_tracks} shows the frequency tracks, $f(t)$, for multiple realizations of the parameters using this procedure. It can be observed that the frequency tracks are uniformly distributed within the region limits for both the dominant and the SASI components.

We consider three cases:
\begin{itemize}
    \item{\it Standard neutrino-driven supernova, no SASI:} These signals correspond typically to the most common case of progenitors ($M>10~M_{\odot}$) in cases where SASI is not present. Only the dominant mode appears and the minimum GW emission is set to $0.4$~s. The parameters of the frequency evolution are chosen to range the values discussed in  Sect.~\ref{sec:FreqEvol}. The boundary of the region of possible frequency evolutions is plotted in red in Fig~\ref{fig:StatSpectrogram}).
    \item{\it Standard neutrino-driven, with SASI:} Same as above but for cases with SASI. In this case, in addition to the dominant mode, a SASI mode is added linearly. Its frequency range spans lower frequencies than the dominant mode. The boundaries of this range are plotted in cyan in Fig~\ref{fig:StatSpectrogram}).
    \item{\it Short neutrino-driven supernovae signals:} These signals correspond typically to low mass progenitors ($M<10~M_{\odot}$) that explode easily and produce very short GW signals. Their duration is limited to the range $0.1-0.4$~s. In this case, only the dominant mode is considered since SASI typically takes longer times to develop.
\end{itemize}

Appendix~\ref{app:examples} displays some examples of waveforms generated with the random waveform generator showing the diversity of waveforms that can be obtained.

%%%%%%%%%%%%%%%%%%
\subsection{Root-mean square strain tests}

\begin{figure*}
    \includegraphics[width=0.46\textwidth]{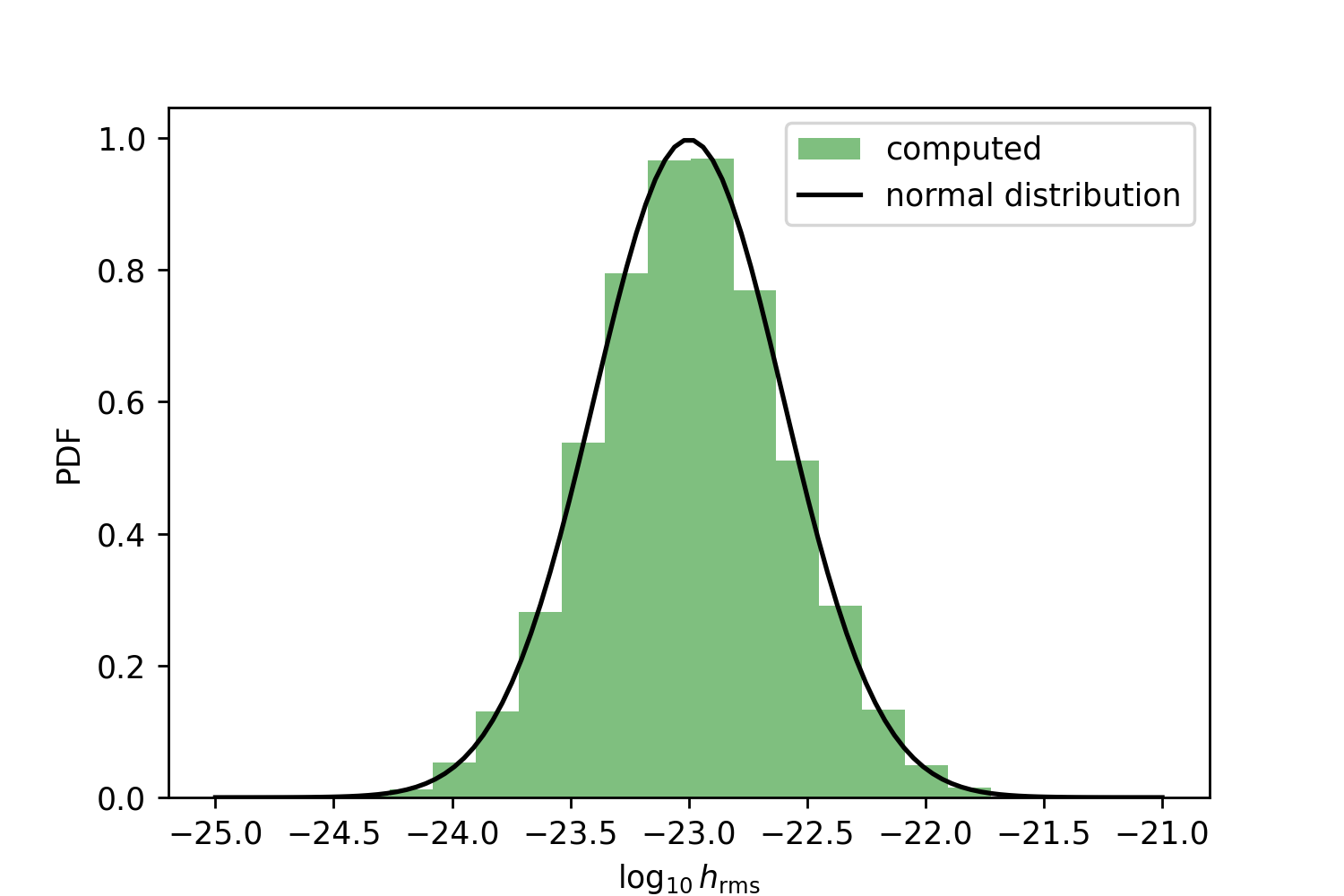}
    \includegraphics[width=0.45\textwidth]{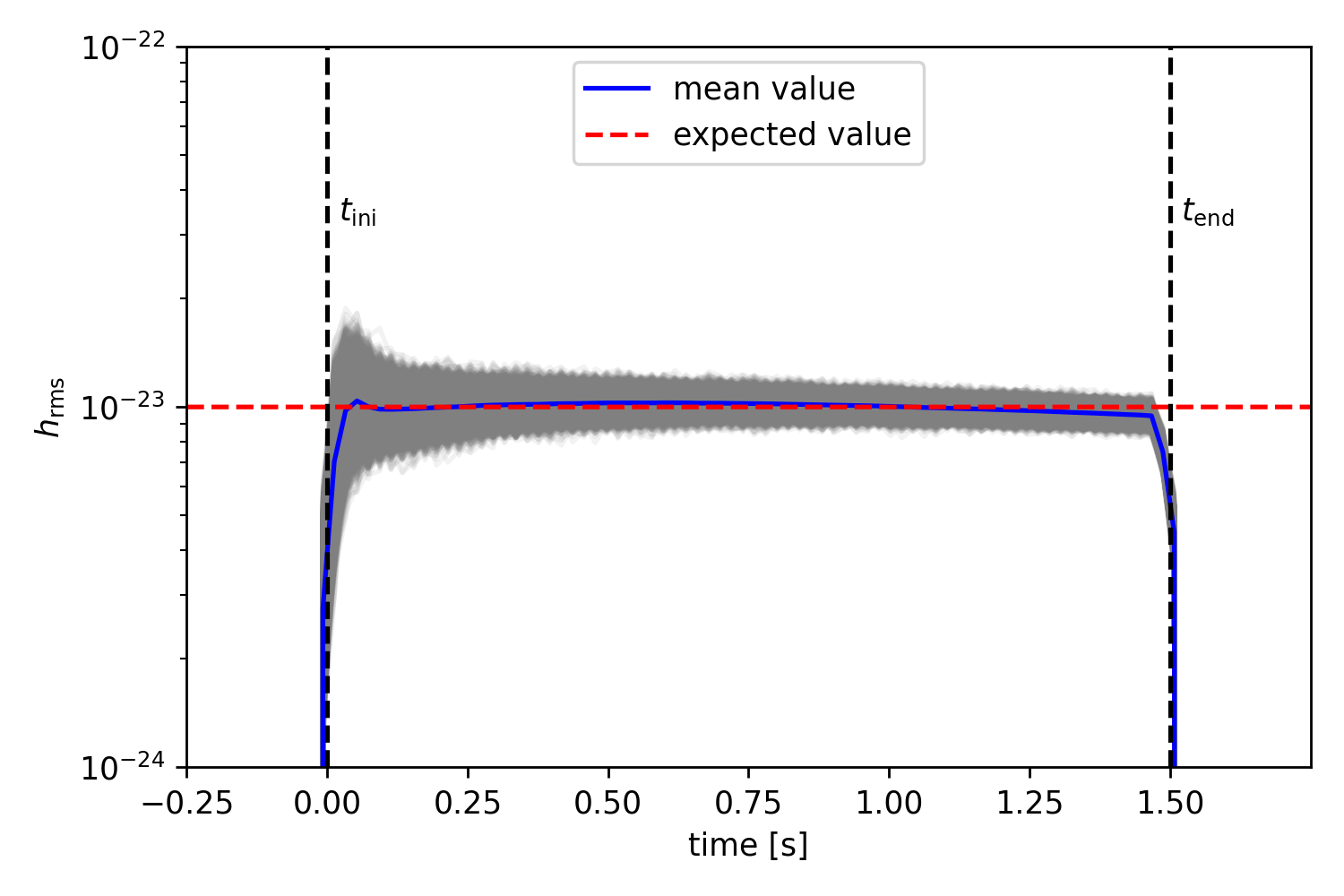}
\caption{Root mean square value of the waveform. The left panel shows a histogram of $log_{10} h_{\rm rms}$ for $10 000$ realizations of a phenomenological waveform at a distance of $10$~kpc. Overplotted (black line) is a normal distribution corresponding to the mean and standard deviation of the waveforms discussed in Section~\ref{sec:par:amplitude}. The right panel shows the time evolution of $h_{\rm rms}$ computed with a sliding window of $50$~ms for $10000$ realizations of a set of waveforms with the same injected rms strain and its mean (blue curve). The red dashed line corresponds to the expected value in the interval $[t_{\rm ini}, t_{\rm end}]$. }
\label{fig:loghrms:test}
\end{figure*}

We have performed several tests to check that the amplitude of the generated waveform follows the selected input parameters. As a measure of the amplitude we use its root-mean-squared value computed as
\begin{equation}
    h_{\rm rms} = \sqrt{<|h|^2>}.
\end{equation}
Note that this quantity is different to $\bar h_{\rm rms}$, which contains an additional angular average, while $h_{\rm rms}$ here is the rms value for a given observation angle. 

By construction, $\log_{10} h_{\rm rms}$ should follow a normal distribution given by the mean and standard deviation given in Eq.~\ref{eq:loghrms}. The left panel of Fig.~\ref{fig:loghrms:test} shows the distribution of $\log_{10} h_{\rm rms}$ for a set of  $10^4$ realizations of the same parameters, except for the random orientation, at $10$~kpc. One can see that it follows very closely the expected normal distribution (black curve). The median and standard deviation of $\log_{10} h_{\rm rms}$ are $-22.9948$ and $0.4005$, respectively, that deviate about $0.02\%$ and $0.1\%$, respectively, from the expected values.

Given that the frequency changes with time, that the global normalization, $h_{\rm rms}$, is correct does not imply immediately that the time evolution of the amplitude is correct as well (in this case it should be constant). To test the time evolution we compute the root-mean-square of the strain using a sliding window of $50$~ms. The right panel of Fig.~\ref{fig:loghrms:test} shows the time evolution of the rms value for  $10^4$ realizations of the same waveform (random orientation) at $10$~kpc, and fixing $\log_{10} \bar h_{\rm rms}=-23.0$, but without any amplitude variability. The differences in rms appear due to the intrinsic stochastic variability of the waveforms and observing angle. However, the mean value of all rms values is constant in time and coincides perfectly with the expected value. 

\begin{figure}
    \includegraphics[width=0.46\textwidth]{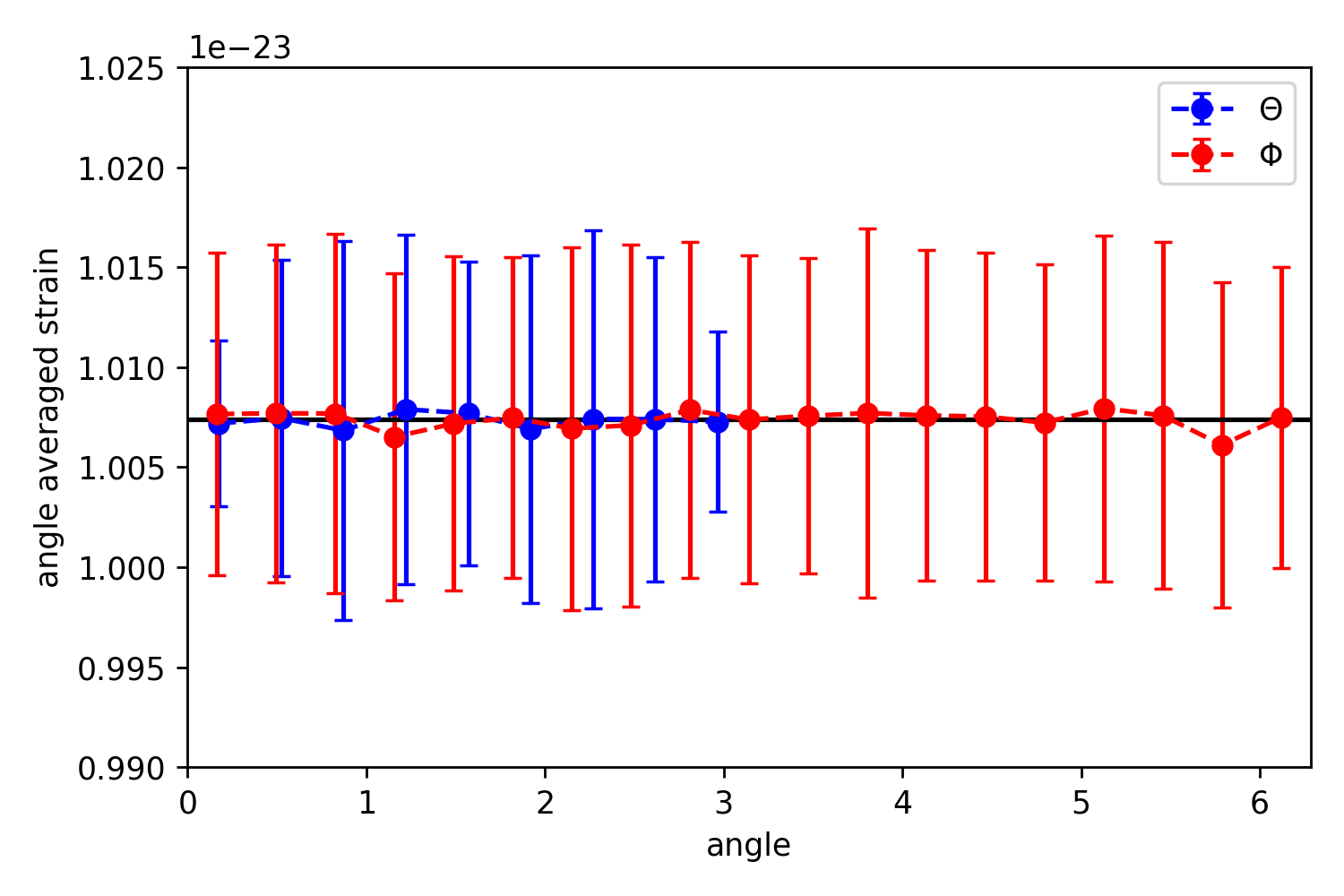}
\caption{Mean rms strain (points) and standard deviation (error bars) for $10^4$ realizations with the same injected rms strain binned according to the observing angles $\Theta$ (red) and $\Phi$ (blue). The black line shows the mean value over all realizations.}
\label{fig:isotropy}
\end{figure}

We have also tested that the source is statistically isotropic. Since the phenomenological waveforms correspond to a non-rotating object, there is no preferred direction for GW emission. The symmetry is broken because of the stochasticity of the perturbations that introduce different perturbations in the different multipoles, which have different angular responses. However, if one averages over a sufficient number of realizations the resulting averaged $h_{\rm rms}$ values should be independent of the observing angles. We have tested this with $10^4$ realizations of the same waveform with the same injected rms strain. To check for isotropy we bin the data over $\Theta$ and $\Phi$ and compute the mean rms strain and standard deviation for all the realizations in each bin. Results are displayed in Fig.~\ref{fig:isotropy}. The variation of the mean rms strain across both observing angles is smaller than typical fluctuations between different realizations with the same angle (characterized by the standard deviation). Therefore, we can conclude that the results are consistent with a set of statistically isotropic sources.

%% file: comparison.tex
%%%%%%%%%%%%%%%%%%%
\section{Comparison with simulations}
\label{sec:comparison}

{A final question that we want to address is the similarity between the waveforms generated by \texttt{ccphen v4} and those from numerical simulations. For this purpose, we compare numerical relativity simulations with ccphen waveforms with similar morphologies.}

{To find phenomenological waveforms that are similar to the ones present in the catalogue, we define an optimization function which represents the squared relative difference between the catalogue waveform and ccphen\footnote{The optimization processed is performed with \texttt{scipy.optimize.minimize} of \texttt{scipy} library \cite{2020SciPy-NMeth}.}:}

\begin{equation}
    R(S_{cat}, S_{ccp}) = \frac{MSE(S_{cat}, S_{ccp})}{h_{rms, cat}},
    \label{e:obfunc}
\end{equation}
{where $S_{cat}$ the spectrogram of a given catalogue waveform, $S_{ccp}$ a spectrogram of a ccphen waveform and $h_{rms, cat}$ the root-mean-square value of the catalogue waveform. If we employ a ccphen model with only dominant mode, we optimize the following parameters:}

\begin{equation}
    R(S_{cat}, S_{ccp}(t_{ini, d}, Q_{0, d}, f_{0, d}, f_{1, d}, f_{2, d})),
    \label{e:fgmode}
\end{equation}
{where $t_{ini, d}$ is the start time of the waveform, $Q_{0, d}$ the damping factor and  $f_{0, d}$, $f_{1, d}$ and $f_{2, d}$ the frequencies of the dominant modes. Here, the end time of the dominant mode $t_{end, d}$ is considered to be the end time of the catalogue waveform. If we use a ccphen model with dominant mode and SASI, we optimize the following parameters:}

\begin{multline}
R(S_{cat}, S_{ccp}(t_{ini, d}, Q_{0, d}, f_{0, d}, f_{1, d}, f_{2, d},\\
t_{ini, s}, t_{end, s}, Q_{0, s}, f_{0, s}, f_{1, s}, f_{2, s}, s)),
\label{e:fsasi}
\end{multline}

\begin{figure*}[t]
\subfloat[\label{fig:gct}{Catalog waveform with dominant mode: \textit{model TM1}}]{%
\includegraphics[width=1\columnwidth]{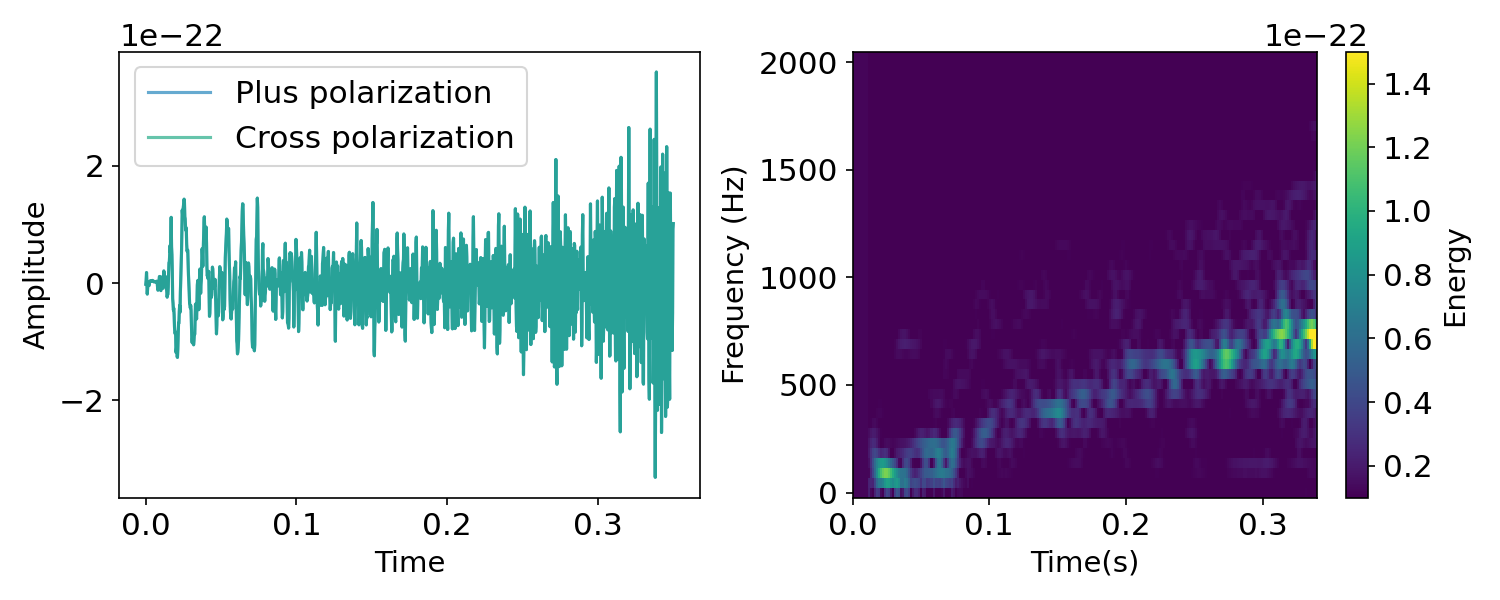}%
}\hfill
\subfloat[\label{fig:sct}{Catalog waveform with dominant mode and SASI: \textit{model mesa20$\_$v$\_$LR}}]{%
\includegraphics[width=1\columnwidth]{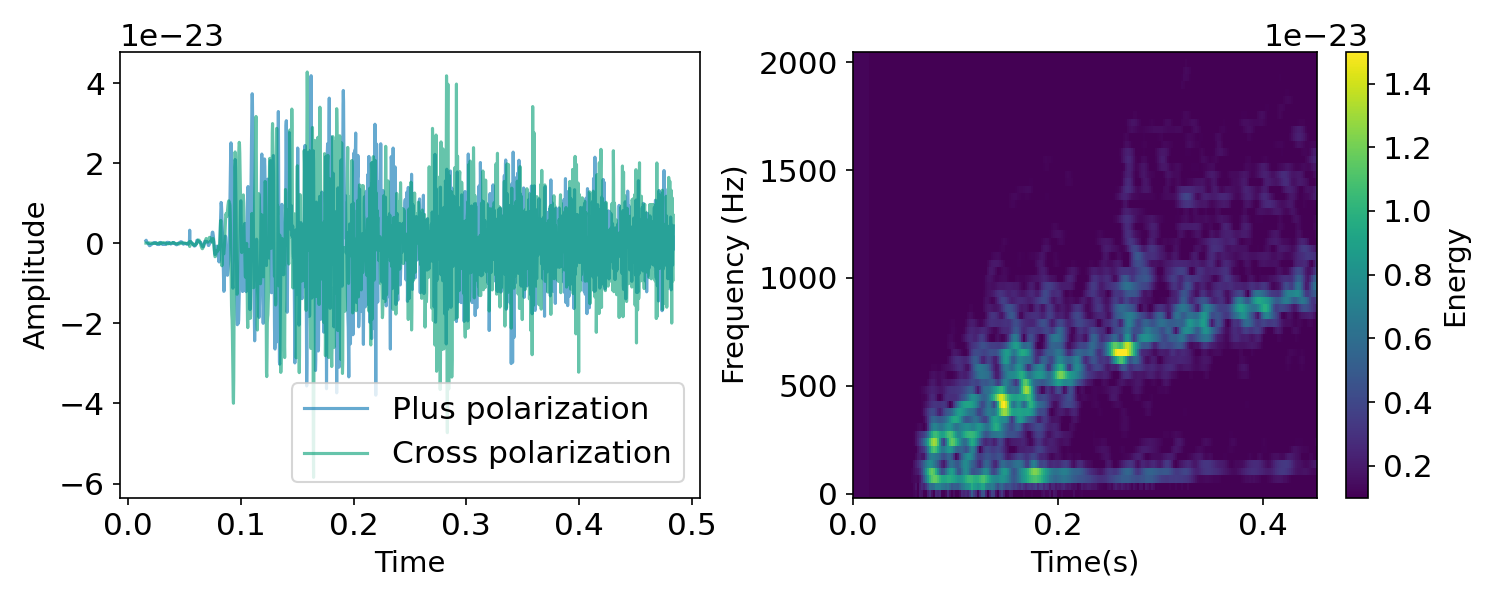}%
}\hfill
\subfloat[\label{fig:gcp}{ccphen waveform with dominant mode}]{%
\includegraphics[width=1\columnwidth]{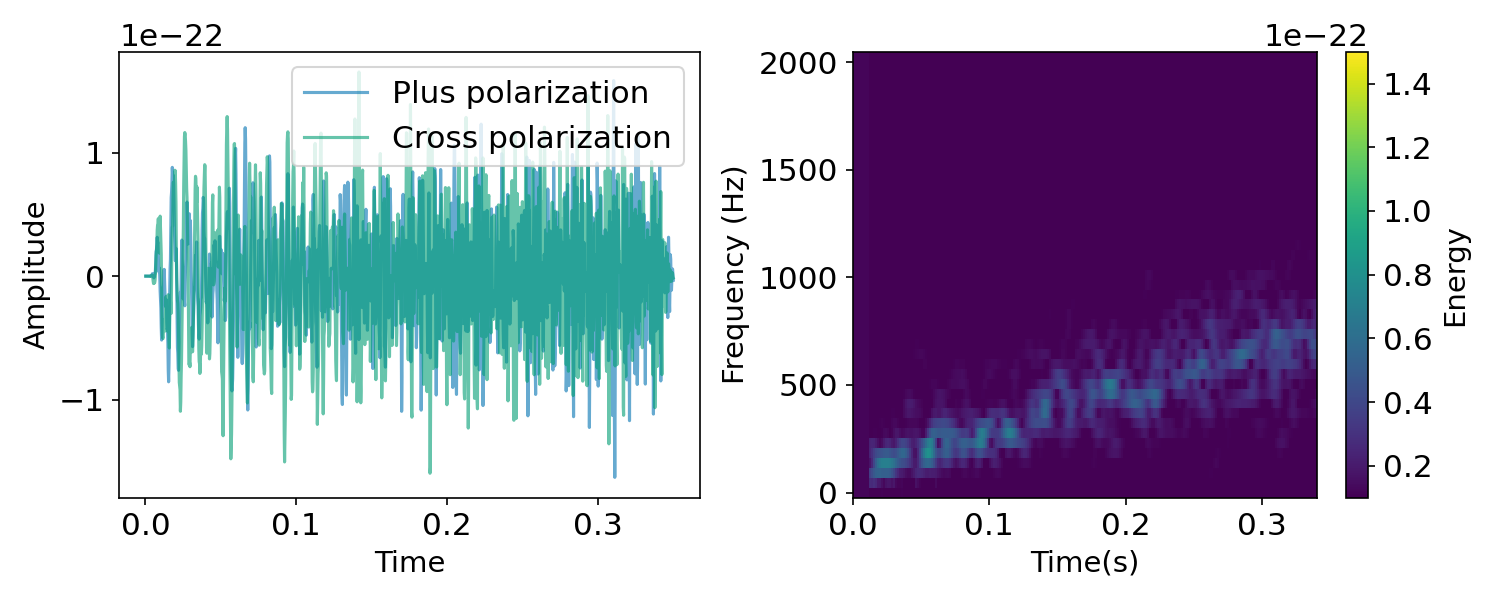}%
}\hfill
\subfloat[\label{fig:scp}{ccphen waveform with dominant mode and SASI}]{%
\includegraphics[width=1\columnwidth]{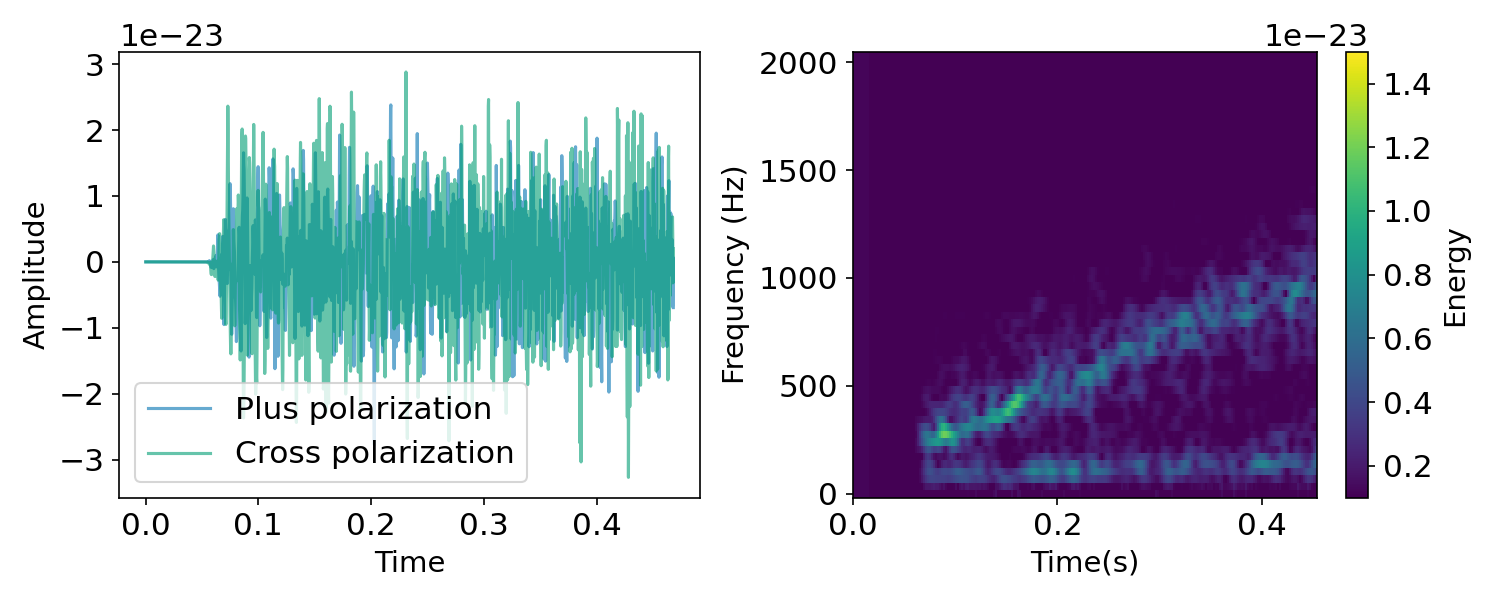}%
}\hfill
\caption{Catalog waveforms (top row) with their optimized phenomenological waveform from ccphen (bottom row) represented in time and time-frequency domain.}
\label{fig:examplectcp}
\end{figure*}
where $t_{ini, s}$ is the start time of SASI, $t_{end, s}$ is the end time of SASI, $Q_{0, s}$ is the SASI damping factor, and $f_{0, s}$, $f_{1, s}$ and $f_{2, s}$ are the SASI frequencies. We can control the amount of SASI with $s$ factor defined in range $[0, 1]$: for $s = 0$ we are essentially using a model with only dominant mode, and for $s=1$  SASI and dominant mode contribution is equal. Note that neither in Eq. \ref{e:fgmode} nor in Eq. \ref{e:fsasi} the amplitude of the ccphen waveform is an optimization parameter since we scale it by $h_{rms, cat}$.

\begin{figure}[t]
 \includegraphics[width=0.48\textwidth]{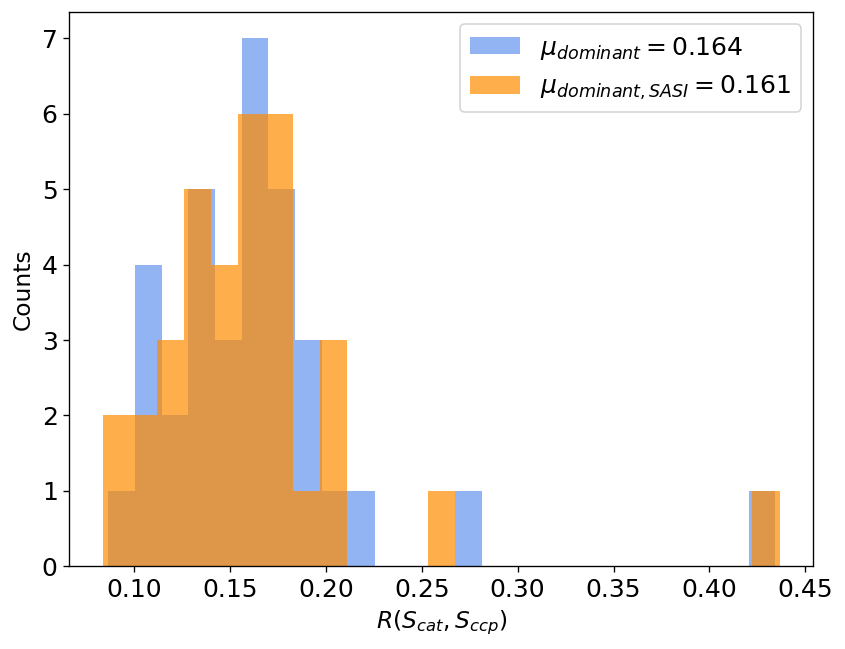}
 \caption{Distribution of the squared relative difference between the catalogue waveforms and optimized ccphen with only dominant mode (blue), and dominant mode and SASI (orange).}
 \label{fig:histOpt}
\end{figure}

{In Fig. \ref{fig:examplectcp} we show the time domain representation and the time-frequency representation of the catalogue waveforms (top row, model TM1 on the left and model mesa20$\_$v$\_$LR on the right), as well as their optimized phenomenological waveforms (bottom row). Their squared relative differences $R(S_{cat}, S_{ccp})$ are 0.1781 for the TM1 model and 0.1572 for mesa20$\_$v$\_$LR model.
On one hand, the phenomenological waveform with the dominant mode in Fig. \ref{fig:gcp} is able to capture the general frequency trend of the model TM1 present in Fig. \ref{fig:gct}, but is not as spread in frequency and it cannot reproduce the final peaks of the catalogue model. On the other hand, the phenomenological waveform with dominant modes and SASI in Fig. \ref{fig:scp} captures the general frequency content of the dominant modes and SASI of model mesa20$\_$v$\_$LR in Fig. \ref{fig:sct}, but it lacks complexity in its frequency components and it does not match the start time of SASI of the catalogue waveform.}

{Fig. \ref{fig:histOpt} shows a histogram summarizing the results applied for all numerical simulations considered in this work. The bulk of the cases differ about $10-20$ points in terms of MSE although some cases may differ as much as $43$ points. Given the stochastic nature of these waveforms, it would be impossible to find a perfect match. This level of accuracy is probably sufficient for many applications, such as the ones discussed in the conclusions.}

%% file: conclusions.tex
\section{Conclusions}
\label{sec:Conclusions}

In this work we present \texttt{ccphen v4}, a new phenomenological waveform generator for core-collapse events, that models the GW strain emitted during these events for the case of non-rotating progenitors, that represent the vast majority of all core-collapse events. The generator is easy to use, with a modern Python interface, and fast, generating $10^3$ waveforms in a few minutes. \texttt{ccphen v4} is a major update of \texttt{ccphen v3} \citep{Lopez2021} including

\begin{enumerate}[i)]
\item polarization, which allows for its use in detector networks, 

\item the use of several waveform components, in particular, the presence of a dominant PNS mode and SASI, and 

\item an improved calibration, taking into account state-of-the-art simulations weighted using Salpeter initial mass function.

\end{enumerate}
The code is publicly available and can be downloaded at  \url{https://www.uv.es/cerdupa/codes/ccphen/}.

The waveforms are parametrized with 12 (7) parameters for models with (without) SASI. The waveforms have a stochastic component, meaning that one can generate multiple realizations of the same set of parameters. At the same time, the use of seeds for the random number generator in the stochastic part ensures that results are reproducible.

We also estimate the bounds of the waveforms parameter space based in a collection waveforms from of state-of-the-art numerical simulations. This allows to generate waveforms automatically within the parameter space of what can be considered realistic. This is of particular interest were a homogeneous coverage of the parameter space is necessary. The generator is easy to modify to incorporate different distributions that may be of interest in different applications.

The waveforms are morphologically similar to those in numerical simulations. The spectrograms of waveforms from numerical simulations typically differ in $10-20$ points from those using \texttt{ccphen} with optimal parameters. This indicates a close similarity of the waveform, which is not expected to have a perfect matching given the stochastic component present. A second test of the similarity is the tests that were performed with the previous version of \texttt{ccphen} \cite{Lopez2021}, in which these waveforms were used to train convolutional neural waveforms to detect CCSN waveforms. The results show that the performance of the CNN when used to detect \texttt{ccphen v3} waveforms was similar when using numerical simulation waveforms. This indicates that at least from the point of view of the CNN, ccphen waveforms are similar to realistic ones. In upcoming work, we will test whether this is the same for \texttt{v4} waveforms.

The main application of \texttt{ccphen v4}, as discussed above is to train CNNs in the context of detection of GWs waves. However, this could be easily extended to regression applications in the context of machine learning. 

Additionally, the waveform generator could be used for testing codes, since it allows for an homogeneous coverage of the parameter space, which allows to assess the performance of a particular detection or parameter estimation pipeline in different parts of the parameter space. This is not currently possible to do with numerical simulation waveforms that are few and cover poorly all physically possibilities. 

The waveforms can also be used to estimate horizon distances for CCSN in current and next-generation detectors. The advantage with respect to current estimations is that it could give you a waveform-independent estimate incorporating all uncertainties of the modelling in a homogeneous way in one single figure, instead of one for each particular waveform. 

Finally, \texttt{ccphen v4} waveforms could be used to establish constraints on non-detected GW events observed by other means, e.g. electromagnetically. This is the case, e.g, of targeted searches \cite{LIGOScientific:2019ryq,Szczepanczyk:2023ihe} in which a simple sine-Gaussian model was used to estimate constraints in GW energy and power. This model could be easily substituted by \texttt{ccphen v4} to improve realism and generality.

%% file: appendix.tex
%%%%%%%%%%%%%%%%%%%%%%%%%%%%%%%%%
\section{Isotropy}
\label{app:isotropy}

Here we estimate the relations that hold between the expected values of  $|I_{2m}|^2$ in the case of a statistically isotropic source. It is convenient to express the mass quadrupole moments as \cite{Raynaud2022}
\begin{eqnarray}
I_{20} &=& -\frac{3}{4}\sqrt{\frac{5}{\pi}} \left(I_{xx} + I_{yy}\right) = \frac{3}{4}\sqrt{\frac{5}{\pi}} I_{zz}  \label{eq:I20}\\
I_{21} &=& -\frac{1}{2} \sqrt{\frac{15}{2\pi}} \left(I_{xz} - i \, I_{yz}\right) \label{eq:I21} \\
I_{22} &=& \frac{1}{4} \sqrt{\frac{15}{2\pi}} \left(I_{xx} - I_{yy} - i \, 2  I_{xy}\right) \label{eq:I22}
\end{eqnarray}
where 
\begin{equation}
I_{ij} \equiv \int \rho({\bf x}, t) \left ( x^i x^j - \frac{1}{3} \delta^{ij} r^2\right) d^3{\bf x}
\end{equation}
are real quantities corresponding to reduced mass quadrupole moment, expressed in a Cartesian coordinates basis.

We consider an ensemble of realizations of a source such that, even if each realization is not spherically symmetric in general, all properties of the ensemble only depend on $r$. The first of properties that are immediate to obtain are
\begin{align}
E[I_{xx}^2] = E[I_{yy}^2] = E[I_{zz}^2],\\
E[I_{xy}^2] = [EI_{xz}^2] = E[I_{yz}^2],\\
E[I_{xx} I_{yy}] = E[I_{xx} I_{zz}] = E[I_{yy} I_{zz}],
\end{align}
which hold because there is no preferred direction in spherical symmetry. We use $E$ to denote the expected value of a quantity in the ensemble.
Appliying these conditions to Eqs.~\eqref{eq:I20}-\eqref{eq:I22}, and computing the expected values of the square of the quadrupolar moments we get 
\begin{eqnarray}
E[|I_{20}|^2] &=& \frac{45}{16 \pi} E[I_{xx}^2], \label{eq:EI20}\\
E[|I_{21}|^2] &=& \frac{15}{4\pi} E[I_{xy}^2], \\
E[|I_{22}|^2] &=& \frac{45}{32 \pi}\left ( E[I_{xx}^2] + \frac{4}{3} E[I_{xy}^2] \right ).\label{eq:EI22}
\end{eqnarray}
The only remaining step is to relate the expectation of $I_{xx}^2$ and $I_{xy}^2$. To simplify this steps of the derivation we consider the fluid as a set of $N$ particles randomly distributed following a spherically-symmetric probability distribution, $P(r)$. The results obtained hereafter can be generalized by taking the continuum limit of the distribution. 

The rest mass density of this collection of particles is
\begin{equation}
\rho({\bf x}) = \sum_{n=1}^{N} \delta ({\bf x} -  {\bf x}_n) m,
\end{equation}
where ${\bf x}_n$ is the location of the nth particle and m is its mass (for simplicity we consider equal mass particles). In this case, the reduce mass quadrupole moment is 
\begin{equation}
I_{ij} = \sum_{n=1}^N \left ( x^i_n x^j_n - \frac{1}{3} \delta^{ij} r^2_n\right) m. 
\end{equation}
Now we can easily compute the expected value of the square of these moments by using the next properties:
\begin{align}
&&E[x] &&= E[y] &&= E[z] &&= 0, \\
&&E[x^2] &&= E [y^2] &&= E[z^2] &&= \frac{1}{3} E[r^2], \\
&&E[x y] &&= E[x z] &&= E[y z] &&= 0, \\
&&E[x^4] &&= E[y^4] &&= E[z^4] &&= \frac{1}{5} E[r^4],\\
&&E[x^2 y^2] &&= E[x^2 z^2] &&= E[y^2 z^2] &&= \frac{1}{15} E[r^4].
\end{align}
These properties are either direct consequence of not having a preferred direction or can be computed by transforming to spherical coordinates and integrating the angular part of the expectation, for which we know the probability is constant. We can also use that the expected value of any of the above quantities for a particle is equal to the one for any other, e.g.
\begin{equation}
E[x_n] = E[x_{n'}] \quad, \quad  \forall \, n,n'. 
\end{equation}
Using these properties we can now compute the next expected values,
\begin{eqnarray}
E[I_{xx}^2] &=& \frac{4 N m^2}{45} E[r^4], \\
E[I_{xy}^2] &=& \frac{3 N m^2}{15} E[r^4], \\
E[I_{xx}I_{yy}] &=& - \frac{2 N m^2}{45} E[r^4].
\end{eqnarray}
which imply that 
\begin{equation}
 E[I_{xx}^2] = \frac{4}{3} E[I_{xy}^2] = -2 E[I_{xx} I_{yy}].   
\end{equation}
Substituting in Eqs.~\eqref{eq:EI20}-\eqref{eq:EI22} we conclude that
\begin{equation}
 E[|I_{20}|^2] = E[|I_{21}|^2] = E[|I_{22}|^2] = \frac{45}{16 \pi} E[I^2_{xx}].
\end{equation}
This result is not surprising, but neither is obvious. 
Finally, we can consider that performing the time average defined in Eq.~\eqref{eq:timeAverage} over a sufficiently long series of data is akin of computing the expected value of an ensemble of realizations so we can identify one with the other and conclude that
\begin{equation}
 <|I_{20}|^2> = <|I_{21}|^2> = <|I_{22}|^2>.
\end{equation}

\section{Waveform examples}
\label{app:examples}

\begin{figure*}[t]
    \centering
    \includegraphics[width=0.325\textwidth]{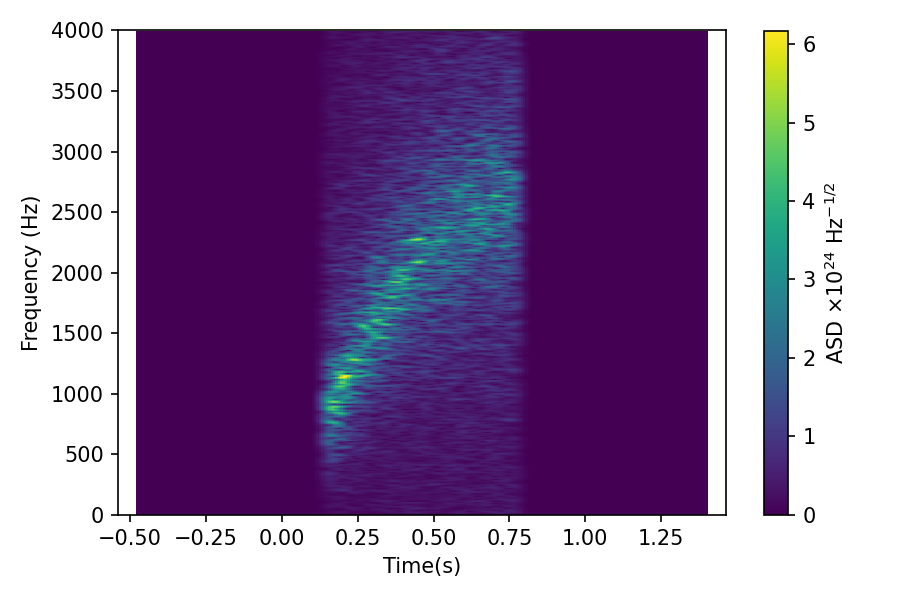}
    \includegraphics[width=0.325\textwidth]{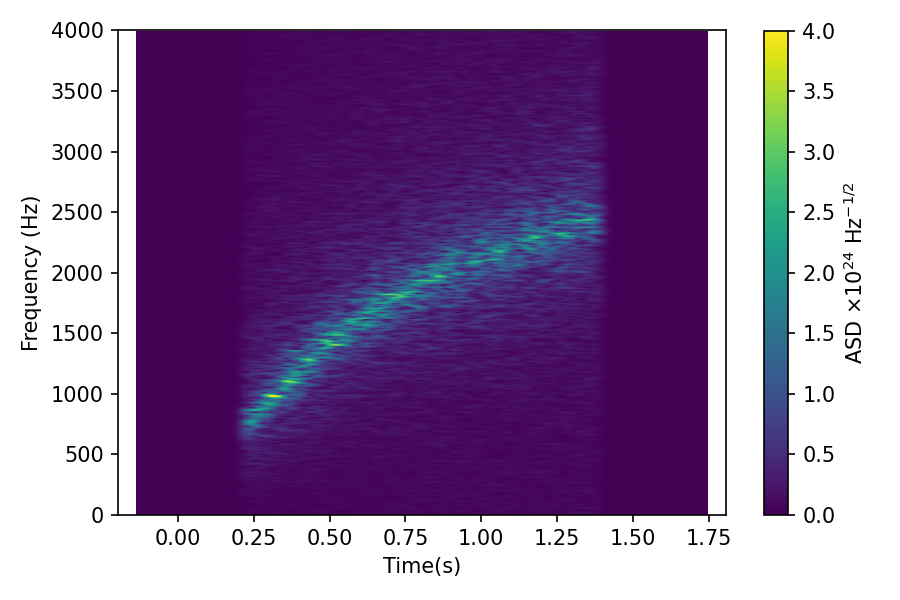}
    \includegraphics[width=0.325\textwidth]{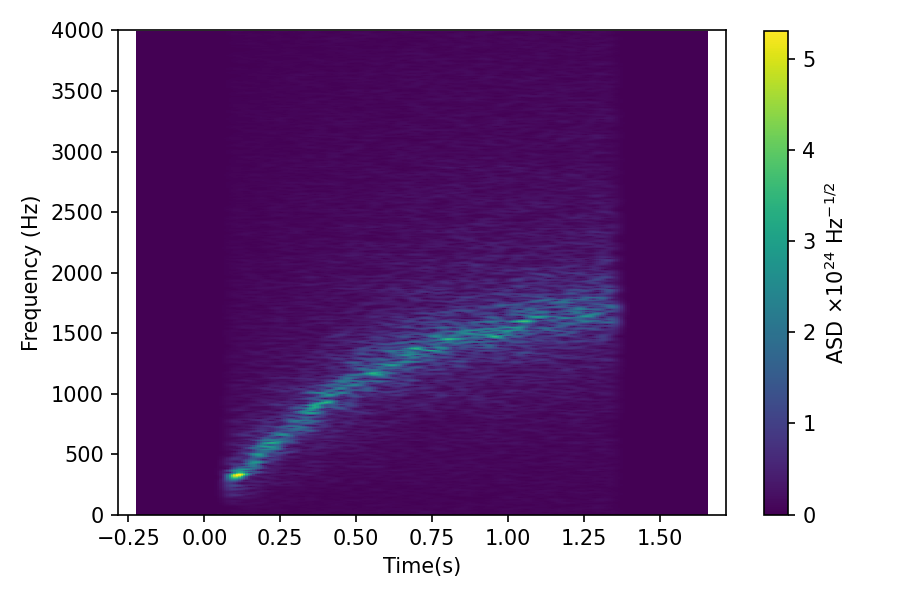}\\
    \includegraphics[width=0.325\textwidth]{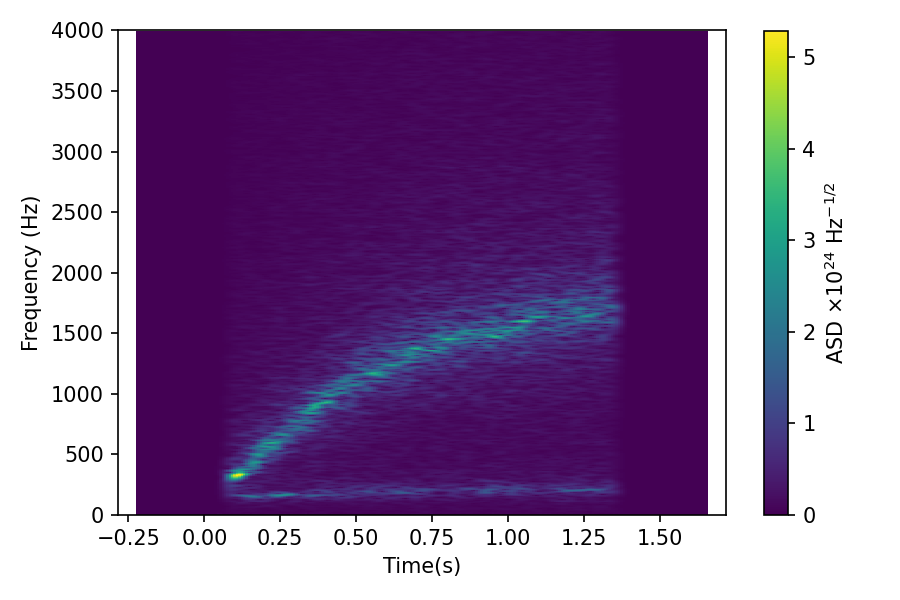}
    \includegraphics[width=0.325\textwidth]{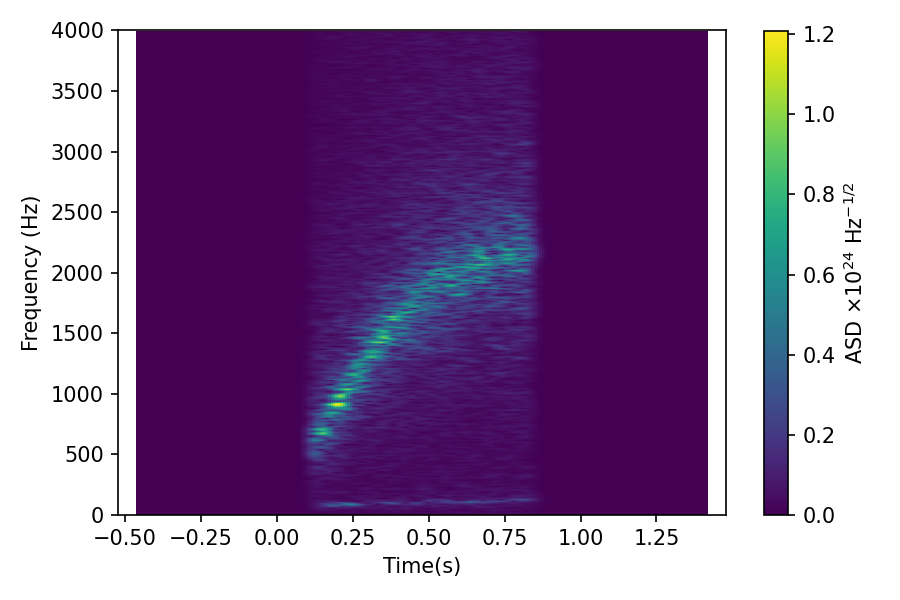}
    \includegraphics[width=0.325\textwidth]{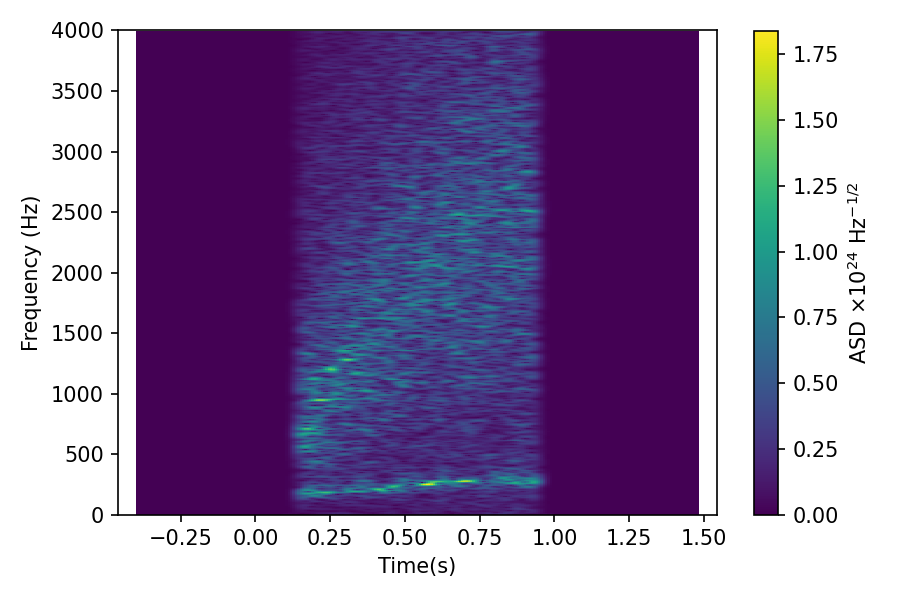}\\
    \includegraphics[width=0.325\textwidth]{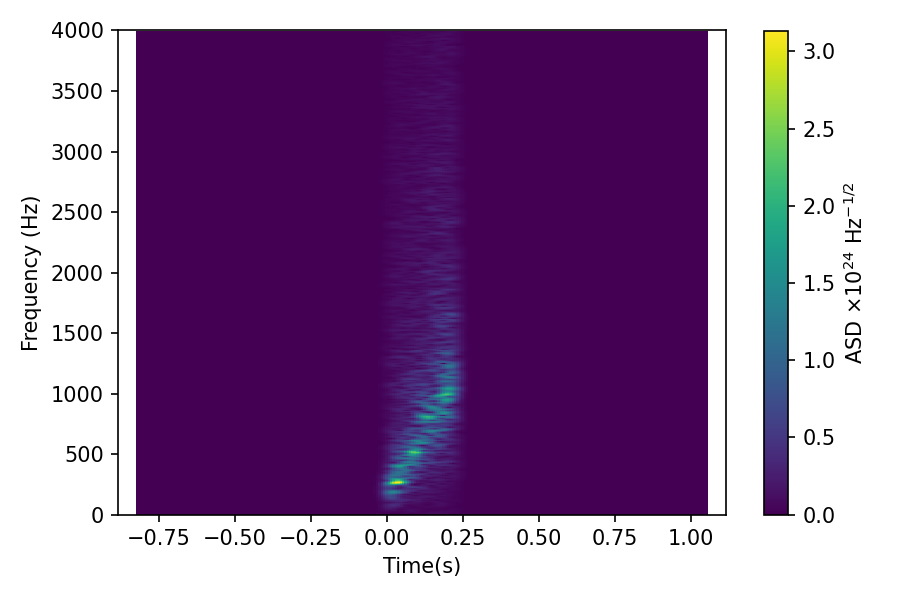}
    \includegraphics[width=0.325\textwidth]{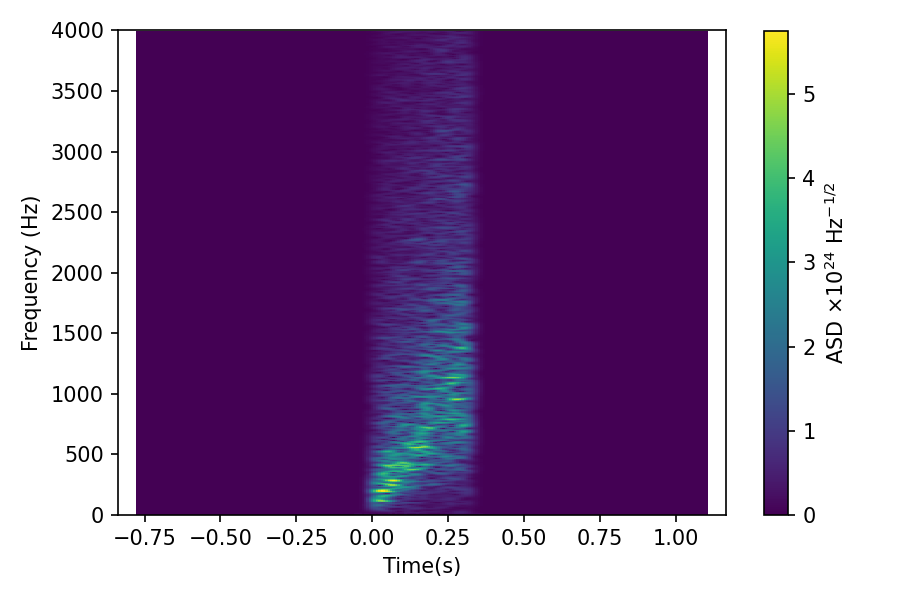}
    \includegraphics[width=0.325\textwidth]{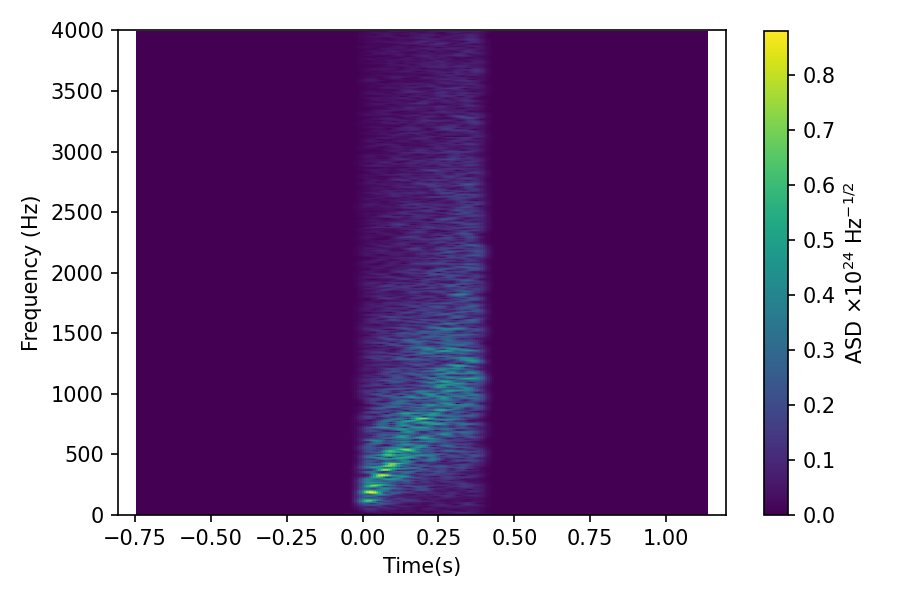}
    \caption{Example of the diversity of the waveforms that can be generated for standard neutrino-driven supernovae without SASI (upper panels), with SASI (middle panels) and for short waveforms (lower panels). For each category we show the spectrograms for three selected examples among the ones produced by the random waveform generator.  }
    \label{fig:spectrogram_examples}
\end{figure*}

Fig.~\ref{fig:spectrogram_examples} shows three selected examples generated by the random waveform generator for each of the three classes described in Section~\ref{sec:randomWfmGen} (9 waveforms in total). Each of the classes is easily recognized by the morphology of the spectrograms. Waveforms containing SASI (middle panels) have a weak track at low frequencies, not present in the other two classes, and short waveforms (lower panels) are characterized by its short duration. Appart from the different basic morphology, there are other differences due to the different parameters used, e.g. the duration, the presence of a delay or the frequency range covered by the different components. The parameter $Q$ affects the spread of the track in the spectrograms. For example, the upper right panel uses a relatively high value $Q=8.25$, which results in a thin track, while the middle right panel uses $Q=1.01$, leading to a very spread dominant mode.

